\documentclass[11pt, a4paper]{article}
\pdfoutput=1

\title{Classical description of black hole degeneracy}

\usepackage[usenames, dvipsnames]{color}
\usepackage[T1]{fontenc}
\usepackage[latin1]{inputenc}
\usepackage{lmodern}
\usepackage{amsmath}
\usepackage[pdftex]{graphicx}
\usepackage{lastpage}
\usepackage{amssymb} 
\usepackage{hyperref}

\usepackage{cite}

\usepackage{esvect}
\renewcommand*\vec{\vv}

\usepackage{xspace}
\newcommand*{\ie}{i.e.\@\xspace}
\newcommand*{\eg}{e.g.\@\xspace}

\newcommand{\ex}{\text{e}}
   
\usepackage{slashed}
\usepackage{braket}
\usepackage{simplewick}
\usepackage{bbm}

\usepackage{xifthen}
\newcommand*\diff{\mathrm{d}} 
\newcommand*\ldiff[2][]{ \ifthenelse{\isempty{#1}}{ \diff #2}{\diff^#1#2} \,} 
\let\limitint\int 
\renewcommand{\int}{\limitint \!} 

\usepackage{subcaption}

\usepackage{titling}
\usepackage{authblk}

\title{Black Hole Evaporation, Quantum Hair and Supertranslations}

\author[1,2]{C\'{e}sar G\'{o}mez\thanks{cesar.gomez@uam.es}}
\author[2,3]{Sebastian Zell\thanks{sebastian.zell@campus.lmu.de}}
\affil[1]{Instituto de F\'{i}sica Te\'orica UAM-CSIC, Universidad Aut\'onoma de Madrid, Cantoblanco, 28049 Madrid, Spain}
\affil[2]{Arnold Sommerfeld Center, Ludwig-Maximilians-Universit\"at, \mbox{Theresienstra\ss e 37, 80333 M\"unchen, Germany}}
\affil[3]{Max-Planck-Institut f\"ur Physik, F\"ohringer Ring 6, 80805 M\"unchen, Germany}

\begin{document}

\allowdisplaybreaks

\maketitle

\vspace{2\baselineskip}
\begin{abstract}
 In a black hole, hair and quantum information retrieval are interrelated phenomena. The existence of any new form of hair necessarily implies the existence of features in the  quantum-mechanically evaporated  radiation.   Therefore, classical supertranslation hair can be only distinguished from global diffeomorphisms if we have access to the interior of the black hole. Indirect information on the interior can only be obtained from the features of the quantum evaporation. 
 We demonstrate that supertranslations $(T^-,T^+) \in BMS_{-}\otimes BMS_{+}$ can be used as bookkeepers of the probability distributions of the emitted quanta  where the first element describes the classical injection of energy and the second one is associated to quantum-mechanical emission. However,  the connection between $T^-$ and $T^+$ is determined by the interior quantum dynamics of the black hole.  We argue that restricting to the diagonal subgroup is only possible for decoupled modes, which do not bring any non-trivial information about the black hole interior and therefore do not constitute physical hair.  It is shown that this is also true for gravitational systems without horizon, for which both injection and emission can be described classically. Moreover, we discuss and clarify the role of infrared physics in purification.
\end{abstract}

\newpage

\tableofcontents

\section{Introduction and Summary}

\subsubsection*{The Puzzle of Black Holes}
A black hole is an extraordinary physical system. While in a classical theory, it is extremely simple for an outside observer, as a consequence of the no-hair-theorem (see \eg \cite{noHair}), its internal quantum complexity measured by the Bekenstein-Hawking-entropy \cite{bekenstein, hawkingRadiation} $N=M^2/M_p^2$ is enormous. Both properties are obviously interrelated. The black hole entropy appears because many different matter configurations can collapse into the same black hole geometry. The no-hair-theorem prevents an outside observer from resolving these differences which remain hidden behind the horizon. Quantum-mechanically, the black hole evaporates \cite{hawkingRadiation} and unitarity requires that along the evaporation process the black hole should deliver back the information which was classically hidden in its interior \cite{page}. This means that although the classical metric has no hair, the evaporation products should have features which compensate for this lack of information. In other words, the quantum radiation emitted during the black hole evaporation should carry the same information which the classical no-hair-theorem prevents us to extract from the geometry. 

In the last twenty years it has become popular to use the AdS/CFT correspondence as strong indication of the unitarity of black hole evaporation. However, this hope will not be fulfilled until counting with the CFT dual of a small evaporating black hole has been achieved. More generally, we shall argue in this note that the solution to the evaporation problem requires to have a microscopic model of the black hole as a quantum system --  whether obtained from AdS/CFT or differently. In \cite{NPortrait,1/NHair}, we have developed such a model that, among other things, indicates the existence of forms of quantum hair effects of order $1/N$. Moreover, and in a model independent way, it is easy to see that taking into account the change of the black hole mass due to Hawking evaporation leads to deviations from featureless emission on precisely this order of magnitude \cite{giaThermal}.

\subsubsection*{Classical BMS-Hair}
Recently, a new way to attack the problem has been suggested \cite {StromingerMemory, HawkingFirst} based on asymptotic BMS-symmetries \cite{BMS}. This approach has received widespread attention (see \eg \cite{BMSotherPenna, BMSotherDieter, BMSotherFlanagan, BMSotherDonnay, BMSotherBlau, BMSotherBarnich, BMSotherArtem, compereMatching, compereFinite, BMSotherEling, BMSotherDonnay2, BMSotherCai}). In particular, a potential new form of classical hair for a black hole has been proposed \cite{HPS1, HPS2}.
The idea is simply the following. One starts with a black hole of mass $M$ and injects an energy $\mu$ in the form of incoming radiation with some angular features.\footnote
{All quantities will be properly defined at the beginning of section \ref{sec:hair}.}
This incoming radiation can be associated with a supertranslation in $BMS_{-}$ which we denote by $T^-$. Classically, the resulting system is a black hole with total mass $M+\mu$ but supertranslated by $T^-$. One can do the same construction with identical $\mu$ but with different angular features, \ie different supertranslations $T_i^-$, to obtain a family of different metrics all of them with the same ADM-charges. Thus, it seems that one can indeed define classical hair if all these metrics sharing the same ADM-charges are physically inequivalent. At the classical level, this means that those metrics are not just the same metric written in different coordinate systems, \ie that they are not related by a globally defined diffeomorphism. 
 As we shall elucidate, the problem with this form of classical hair is that for an observer outside, there is no way to decide if all these metrics are different or simply the same metric in different coordinates. In order to decide that, the observer needs to have information about the interior of the black hole. In summary, defining hair by means of the classical gravitational memory associated to some incoming radiation is only operative if somehow we can have extra information about the memory effects in the interior of the black hole which is, in a different guise, the essence of the no-hair-theorem.

\begin{figure}	
	\begin{subfigure}{\textwidth}
		\begin{center}
			\includegraphics[width=0.55\textwidth, trim={0cm 20.7cm 1.5cm 4cm},clip]{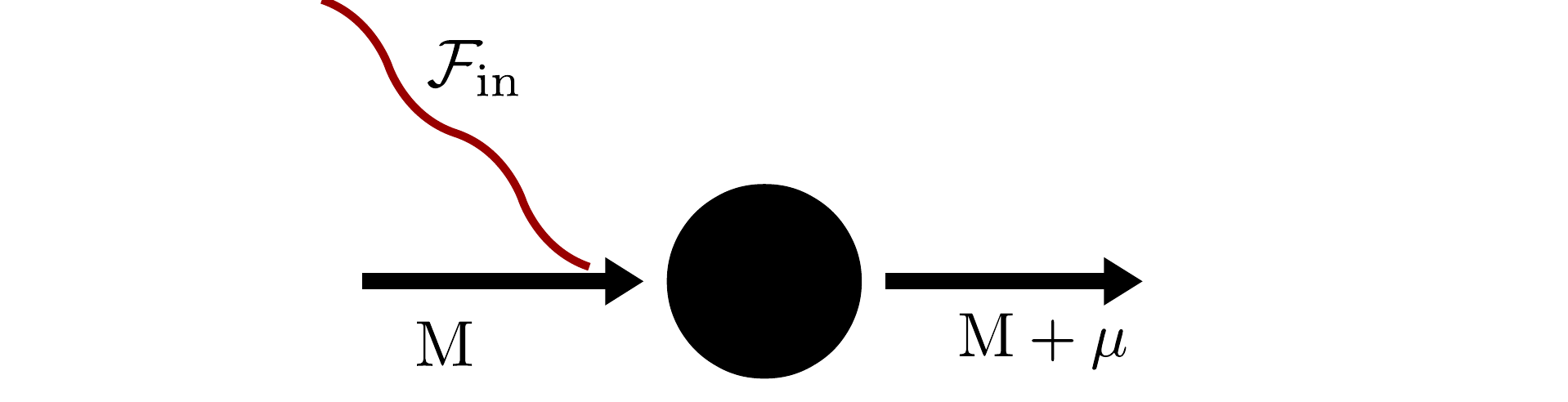}
			\caption{In the classical limit, the change of the black hole only depends on the total absorbed energy $\mu$, in line with the no-hair-theorem. The black hole cannot emit.}
		\end{center}
	\end{subfigure}
	\begin{subfigure}{\textwidth}
		\begin{center}
			\includegraphics[width=0.55\textwidth, trim={0cm 20.7cm 1.5cm 3.5cm},clip]{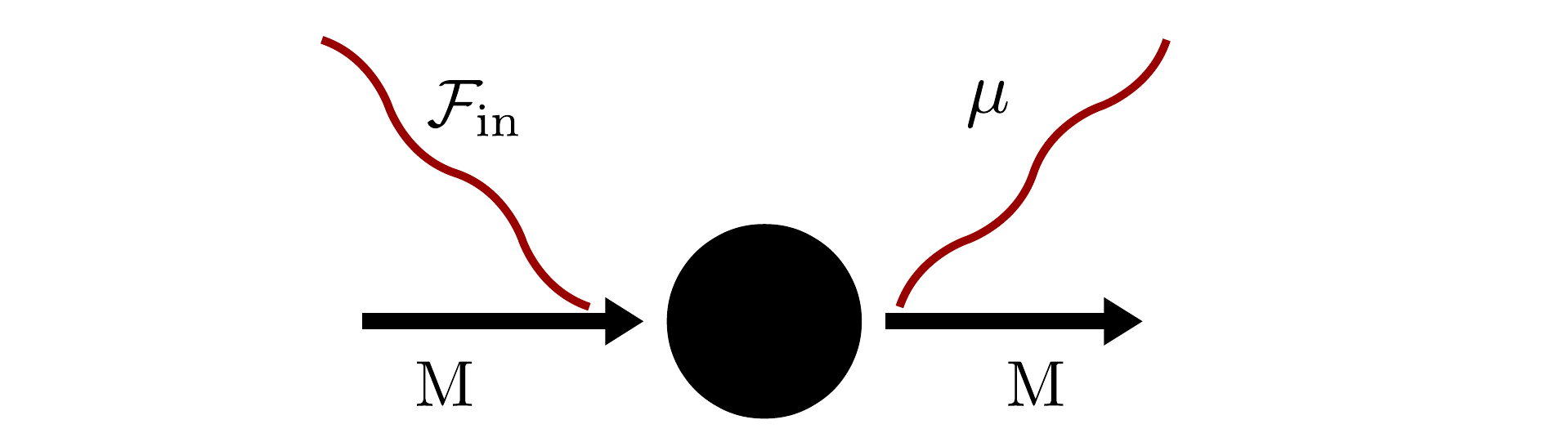}
			\caption{Also in the semi-classical limit, the change of the black hole only depends on the total absorbed energy $\mu$. The black hole can evaporate, but the evaporation products are featureless. In particular, they are distributed isotropically.}
		\end{center}
	\end{subfigure}
	\begin{subfigure}{\textwidth}
		\begin{center}
			\includegraphics[width=0.55\textwidth, trim={0cm 20.2cm 0cm 3.5cm},clip]{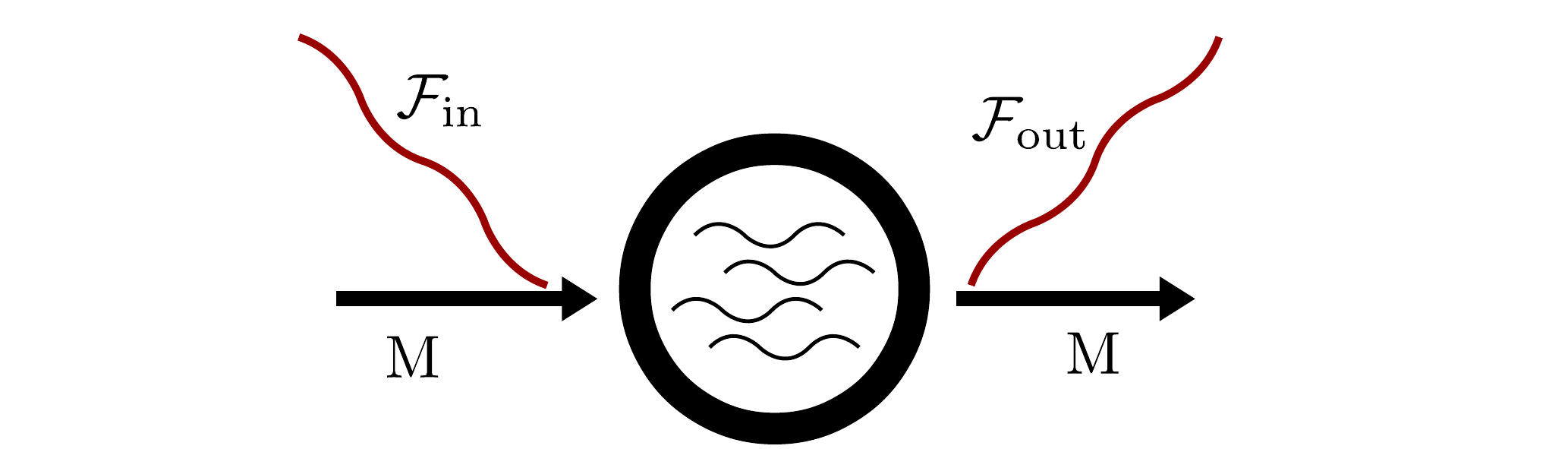}
			\caption{In the fully quantum treatment, the incoming radiation $\mathcal{F}_{\text{in}}$ interacts with the microscopic description of the black hole. In doing so, it changes the microstate of the black hole so that it can emit radiation $\mathcal{F}_{\text{out}}$ with non-trivial angular profile.}
		\end{center}
	\end{subfigure}	
	\caption{Absorption of a wave with energy profile $\mathcal{F}_{\text{in}}$ by a black hole of mass $M$ and possible subsequent evaporation in the classical, semi-classical and fully quantum treatment.}
	\label{fig:limits}
\end{figure}

Fortunately, there is an indirect way to decide from the outside whether two black hole metrics defined by injecting the same amount of energy but with different angular features are physically different or not. We can just wait until the black holes emit some radiation and compare the radiation produced by the two black holes. For simplicity, we restrict ourselves in our discussion to a pure theory of gravity in which only gravitational radiation can be emitted. The corresponding process is depicted in figure \ref{fig:limits}, where we distinguish the classical, semi-classical and purely quantum contributions. The first thing to be noticed is that this test is purely quantum in the sense that only quantum-mechanically, the black hole can emit radiation. The second thing is that the information we can get on the emitted quantum radiation by actual measurements is necessarily encoded in the form of probability distributions. Thus, if those black holes defined by different $T^i_{-}$ are indeed different, we should expect that the corresponding quantum probability distributions are also different. 
\renewcommand{\floatpagefraction}{.8}

\subsubsection*{Insufficiency of Calculation in Classical Background Metric}
  It is natural to expect that this difference has a non-trivial projection on deviations from isotropy, \ie that the emitted quanta carry angular features. Then we obtain quantum probability distributions $P_{i}(\theta,\phi)$ from the measurement of the radiated quanta. We can use those to define a classical supertranslation $T_i^+$. By that we simply mean a classical supertranslation with the flow of emitted radiation determined by the measured quantum probability functions $P_{i}(\theta,\phi)$. In this sense, the former experiment produces a set of couples $(T_i^-, T_i^+)$ where the first supertranslation in $BMS_-$ is classical and the second one in $BMS_+$ is determined by the quantum probability distribution. From this point of view, if the classical $T_i^-$ really implants hair, then the quantum $T_i^+$ should be non-trivial, \ie contain spherical harmonics with $l\geq 2$.  The crucial point is that this behavior cannot be achieved by the standard Hawking computation performed in a supertranslated Schwarzschild metric, \ie as pair creation in the background vacua defined by the near horizon geometry. The reason is that the supertranslation acts as a diffeomorphism near the horizon and does not change the local geometry. Therefore, it does not suffice if the $P_i$ only depend on the injected radiation and the geometry of the black hole. Instead, they must also depend on its internal dynamics. 

We can make the argument a bit more quantitative and assume that from the whole energy $\mu$ injected a fraction $\tilde \mu$ is associated to angular features. This means that the part of the incoming classical flow $\mathcal{F}_{\text{in}}$ with angular labels $l\geq 2$ contributes to $\int \ldiff[2]{\Omega} \mathcal{F}_{\text{in}}$ with a value equal to $\tilde \mu$.   Clearly, $\tilde \mu=0$ would correspond to the injection of featureless radiation.\footnote
{In this case, the associated supertranslation will not support angular features and will only project on the $l=0,1$ spherical harmonics.}
 In order to parametrize how the $P_i$ depend on $\tilde{\mu}$ and the internal structure of the black hole, we shall use the typical number of quantum constituents of the black hole. In this sense, we expect $P_i(\theta,\phi; \tilde{\mu}, N)$ where the label $i$ refers to the dependence on the incoming $T_i^-$ and where we identify the number of quantum constituents of the black hole with the entropy $N$. 

Then the natural dimensionless parameter measuring the dependence of $P_i$ on the internal structure is $\tilde{\mu}/\sqrt{N}$ where $\sqrt{N}$ is the black hole mass in Planck units. In this setup, angular features in the evaporation, \ie finite $N$ effects in $P_i(\theta,\phi; \tilde \mu, N)$, depend on the black hole microscopic model. Those will define a couple $(T_i^-, T_i^+)$  generically not in the diagonal subgroup $BMS_0$ of $BMS_- \otimes BMS_+$.  Thus, the first important message of our note will be that the information about $T_i^-$ cannot determine the quantum probability distribution $T_i^+$, \ie we cannot predict the quantum probability distribution solely from the incoming radiation implanting the hair. 

\subsubsection*{Subleading Soft Modes}
 We can investigate how this situation changes in the semi-classical limit $M\rightarrow \infty$, \ie $N\rightarrow \infty$, in which the Hawking computation becomes exact. In this case, the energy associated to features becomes zero so that angular features can only be encoded in zero-energy modes. The effective decoupling of these modes will lead to a $P_i$ identical to the incoming radiation. This produces couples $(T_i^-, T_i^+)$ in the diagonal subgroup $BMS_0$. In more concrete terms, the $\lim_{N\rightarrow\infty} P_{i}(\theta,\phi; \tilde{\mu}, N)$ will only capture local horizon physics or zero-energy modes.\footnote
 {It is important to stress that the pseudo Goldstone-Bogoliubov modes identified in \cite{criticalGas} are not equivalent to near horizon diffeomorphisms and consequently are good microscopic candidates to describe the low energy effective changes of the microstate of the black hole during the process of absorption and evaporation.}

This brings us to our second point, namely how the actual features of the quantum probability distribution $P_{i}$ depend on infrared physics.\footnote
{See \cite{stromingerNew} for a recent suggestion for purification by infrared modes.} 
 We know that in gapless theories such as gravity, evaporation interpreted as a $S$-matrix process has a zero probability amplitude without any accompanying soft gravitons. In order to obtain a finite answer, one has to include the emission of a certain class of soft radiation, namely IR-modes. However, this fact by no means implies that this companion radiation should carry the angular features that we need to purify the evaporation.  On the contrary, we know from infrared physics that IR-radiation is only sensitive to the initial and final scattering states. It is independent of the details of the process or in our case of the microscopic details of the black hole, \ie cannot resolve the microstate.

 Independently of the question to what extent IR-radiation and hard quanta are correlated, we can quantitatively estimate the amount of information we could lose when we integrate over unresolved IR-modes. From well-known results of infrared physics it follows that their number only grows logarithmically with the resolution scale $\epsilon$, \ie  $n_{\text{soft}} \sim -\ln \epsilon$. However, what we have discussed implies that the natural resolution scale of features should be $\epsilon \sim 1/N$. Thus, the second important message of our note is that unresolved IR-modes cannot account for the  bulk of information in black hole evaporation, but could only contribute as a subleading logarithmic correction. The part which carries features is the part of the radiation that can be resolved and that depends not on the infrared divergences but on the inner structure of the black hole, or in scattering language, on the details of the scattering process. 
 A possible candidate is soft non-IR radiation, which is independent of infrared divergences. As it should be, non-IR radiation depends on the details of the scattering process so that it cannot be predicted without a microscopic theory of the black hole. 

\subsubsection*{Summary and Outline}
In summary, non-trivial hair can be only defined by couples $(T_i^-, T_i^+) \in BMS_{-}\otimes BMS_{+}$ where the element in $BMS_{-}$ is {\it classical } and carries some finite energy and the element in $BMS_{+}$ is defined as a bookkeeper of the {\it quantum} probability distribution of the radiated quanta. What concrete element $T_i^+$ is associated with a given $T_i^-$ cannot be derived solely from the classical geometry, but depends on the internal quantum structure of the black hole.  This non-trivial mapping is precisely what makes the quantum hair informative. For a system with non-trivial dynamics, it is therefore impossible to restrict to a subgroup of $BMS_- \otimes BMS_+$.
Predictivity on this quantum output can be only achieved in the zero-energy (or equivalently $N=\infty$ limit) where we only get elements in the diagonal subgroup $BMS_0$.\footnote
{In \cite{HPS1} and \cite{HPS2} it is suggested to constraint the potential values of $T_i^+$ using an infinite set of conserved charges. Imposing these conservation laws makes the corresponding $S$-matrix completely insensitive to the internal structure of the black hole and consequently, in the language we are using here, can only capture unobservable zero-modes.}
But since the soft modes are decoupled once the infrared divergences of the theory are properly taken into account \cite{porrati1, sever, raoul, mischa, porrati2},  they cannot lead to observable features.\footnote
{This decoupling of soft modes is a quantum effect that should not be confused with the existence, for instance in asymptotically Minkowski space time, of a non-trivial family of asymptotically flat connections defining a representation of the $BMS$-group (see \cite{ashtekar} and references therein). This multiplicity of classical inequivalent vacua is quantum-mechanically reabsorbed in the cancellation of infrared divergences.} 

 The outline of the paper is as follows. In section \ref{sec:hair}, we first recap some properties of BMS-gauge. In particular, we show how angular features of radiation define a supertranslation, which can be measured as a memory effect. Moreover, we discuss the role of soft modes. Then we use a combination of injected and emitted radiation of the same total energy to define Goldstone supertranslations as element $(T^-,\,  T^+) \in BMS_{-}\otimes BMS_{+}$. 
In section \ref{sec:implanting}, we first concentrate on a gravitational system without horizon, which we shall call planet for concreteness, and show how we can use Goldstone supertranslations to change its angular distribution of mass. In doing so, the key point is that it is impossible to infer $T^+$ from $T^-$ unless one knows the internal dynamics of the planet. Moreover, we highlight the importance of angular features by showing that it is impossible to determine the angular mass distribution of the planet without access to its interior. Subsequently, we apply Goldstone supertranslations to a black hole. We demonstrate how supertranslations can be used as bookkeeping tool for the emitted quanta. However, without knowledge of the microscopic dynamics of the black hole, they have no predictive power.  We also point out how we can use Page's time to estimate the magnitude of deviations from featureless evaporation.
After concluding in section \ref{sec:conclusion}, we provide a more detailed discussion of IR-physics in appendix \ref{app:IR}. In appendix \ref{app:matching}, we discuss the matching of the supertranslation field in advanced and retarded coordinates and finally we explicitly calculate a Goldstone supertranslation of a planet in appendix \ref{app:planetSolution}.

\section{Quantum Hair}
\label{sec:hair}
\subsection{Recap of BMS-Gauge and Memory Effect}

\subsubsection*{Retarded Coordinates}
We first recap some properties of BMS-gauge, which is defined by the four gauge conditions \cite{BMS}
\begin{equation}
	g_{11} = g_{1A} = 0 \,, \ \ \ \det g_{AB} = r^2 \sin^2\theta \,,
	\label{gaugeConditions}
\end{equation} 
where $A,B, \ldots = 2,3$. Typically, BMS-gauge is used to study a spacetime asymptotically, \ie for $r\rightarrow \infty$, but it is possible to extend the metric to the bulk by imposing the conditions \eqref{gaugeConditions} to all orders in $1/r$. In a typical situation, however, a metric in BMS-gauge does not cover the whole spacetime.

 A metric in BMS-gauge exists both in retarded time $u$, which is suited to describe outgoing radiation, and in advanced time $v$, which is suited to describe incoming radiation. The matching between these two metrics will be crucial for our treatment. Explicitly, an asymptotically flat metric in retarded time takes the form \cite{BMS}:
\begin{align}
\diff s^2=& \left(-1 + \frac{m^+_B}{r} + O(r^{-2})\right)\diff u^2 - \left(2+O(r^{-2})\right) \diff u \diff r \\
&+ r^2 	\left(\gamma_{AB} + C_{AB}^+ r^{-1} + O(r^{-2})\right) \diff x^A \diff x^B + O(r^{-2}) \diff x^A \diff u\,, \label{BMSMetricU}
\end{align}
where the metric on the sphere has to fulfill the requirement $\det g_{AB} = r^2\sin^2 \theta$. Here $m_B^+$ is the Bondi mass, $\gamma_{AB}$ the standard metric on the sphere and 
\begin{equation}
	C^+_{AB} = \left(2 D_A D_B - \gamma_{AB}D^2 \right)C^+
	\label{supertranslationField}
\end{equation}
is determined by the supertranslation field $C^+$, where $D_A$ is the covariant derivative on the sphere. It is helpful to expand the supertranslation field in spherical harmonics. Then the mode $l=0$ represents a time shift and the mode $l=1$ corresponds to spatial translations. Therefore, all modes with $l\geq 2$ define proper supertranslations.  Metrics with different values of $C^+$ are connected via asymptotic diffeomorphisms, \ie the choice of the supertranslation field constitutes a residual gauge freedom of BMS-gauge. These diffeomorphisms are the famous supertranslations. Therefore, we can define a supertranslation $T^+$ by the change it induces in the supertranslation field:
\begin{equation}
	T^+ := \Delta C^+ \,.
	\label{supertranslation}
\end{equation}

In order to analyze the effect of supertranslations, we will need the constraint equation $G_{00}=8\pi G T_{00}$, whose leading order reads in BMS-gauge:
\begin{equation}
\partial_u m^+_B =  \frac{1}{4G}D^2(D^2+2) \partial_u C^+ -\mathcal{F}_{\text{out}} \,,
\label{fullConstraintU}
\end{equation}
where 
\begin{equation}
	\mathcal{F}_{\text{out}} = \frac{1}{8}(\partial_u C^+_{AB})(\partial_u C^{+AB}) + 4\pi \lim_{r\rightarrow \infty} (r^2T_{uu})
	\label{outgoingEnergy}
\end{equation}
is the total incoming null energy, composed of gravitational waves (first summand) and other forms of gravitating energy (second summand). 

\subsubsection*{Advanced Coordinates}
The situation in advanced coordinates is very similar. The metric takes the form
\begin{align}
\diff s^2=& \left(-1 + \frac{m^-_B}{r} + O(r^{-2})\right)\diff v^2 + \left(2+O(r^{-2})\right) \diff v \diff r \\
&+ r^2 	\left(\gamma_{AB} + C_{AB}^- r^{-1} + O(r^{-2})\right) \diff x^A \diff x^B + O(r^{-2}) \diff x^A \diff v\,, \label{BMSMetricV}
\end{align}
where the supertranslation field and the supertranslations in advanced coordinates are defined as in \eqref{supertranslationField} and \eqref{supertranslation}. The constraint equation becomes
\begin{equation}
\partial_u m^-_B =  \frac{1}{4G}D^2(D^2+2) \partial_u C^- +\mathcal{F}_{\text{in}} \,,
\label{fullConstraintV}
\end{equation}
where $\mathcal{F}_{\text{in}}$ is the incoming energy, in analogy to \eqref{outgoingEnergy}.

\subsubsection*{Measurement of the Supertranslation Field: Memory Effect}
As already discussed, one can change the value of the supertranslation field by a diffeomorphism. Therefore, it follows by general covariance that the value of the supertranslation field cannot have in general any experimental implication. However,  since it corresponds to physical outgoing or ingoing radiation, the difference of the supertranslation field at different times does have experimental implications: It describes the memory effect caused by the radiation, \ie a permanent displacement of test masses after the radiation has passed \cite{memoryFirst, StromingerMemory, wald}. 

We will restrict ourselves to a simple situation in which we start with some stationary metric $g^1_{\mu\nu}$ and we finish in a different stationary metric $g^2_{\mu\nu}$. In between, there is a radiation epoch, \ie $\mathcal{F}_{\text{in/out}}$ only has support during this time span. Asymptotically on $\mathcal{J}^\pm$, the process defines a non-stationary metric interpolating between $g^1_{\mu\nu}$ and $g^2_{\mu\nu}$ which should be a solution to the Einstein equations.

 Since Birkhoff's theorem implies that we can set $\partial_A m_B^\pm  = 0$ in a stationary metric, we can single out the zero-mode from \eqref{fullConstraintU} by integrating over the sphere:
\begin{equation}
  \mu^+ = -\frac{\int \ldiff{u} \int \ldiff[2]{\Omega} \mathcal{F}_{\text{out}}}{4\pi} \,,
  \label{deltaMU}
\end{equation}
where we first consider retarded time and $ \mu^+ = m^+_{B,\,2}- m^+_{B,\,1}$ is the total change of Bondi mass due to the radiation epoch. This formula shows explicitly that the Bondi mass $m^+_B$ is monotonically decreasing, \ie it measures the energy which has not yet left the bulk.  Defining the emitted energy with non-trivial angular distribution as $\Delta \tilde{\mathcal{F}}_{\text{out}} := \int \ldiff{u}\mathcal{F}_{\text{out}} - \mu^+$, the constraint \eqref{fullConstraintU} becomes
\begin{equation}
0 =  \frac{1}{4G}D^2(D^2+2) T^+ - \Delta \tilde{\mathcal{F}}_{\text{out}} \,.
\label{constraintU}
\end{equation}
 Thus, angular features in the outgoing radiation induces a supertranslation $T^+=\Delta C^+$. Note that it is independent of the total emitted energy $\mu^+$.

 In advanced coordinates, we get from the constraint \eqref{fullConstraintV}:
\begin{equation}
\mu^- = \frac{\int \ldiff{v} \int \ldiff[2]{\Omega} \mathcal{F}_{\text{in}}}{4\pi} \,.
\label{deltaMV}
\end{equation} 
The advanced Bondi mass $m^-_B $ is monotonically increasing, \ie it measures the energy which has already entered the bulk. Defining $\Delta \tilde{\mathcal{F}}_{\text{in}} := \int \ldiff{v}\mathcal{F}_{\text{in}} - \mu^-$, the constraint \eqref{fullConstraintV} becomes
\begin{equation}
0 =  \frac{1}{4G}D^2(D^2+2) T^- + \Delta \tilde{\mathcal{F}}_{\text{in}} \,.
\label{constraintV}
\end{equation}
 This formula implies that an advanced supertranslation $T^-$ tracks angular features in the incoming radiation.

\subsection{Goldstone Supertranslations}
As already pointed out, we shall define hair on the basis of scattering processes where some injected gravitational energy is radiated back by the system. The hair will be encoded in the angular features of the injected radiation and the outgoing radiation. In this sense, we define hair as a typical {\it response function}. Through these formal scattering processes we define a map relating gravitational systems, black holes or planets, in different states sharing the same values for all the ADM-conserved quantities. We denote this induced map a Goldstone supertranslation since it relates states which are degenerate in energy. Note that this scattering definition of hair is tied to the mechanism of radiation whatever it could be. 

\subsubsection*{Relationship to Antipodal Matching}
 As a first step, it is important to discuss whether there are general constraints on this scattering process. Namely, it has been suggested in \cite{StromingerBMSInvariance, stromingerLecture} that any gravitational $S$-matrix in an asymptotically flat spacetime must satisfy the following relation for an arbitrary initial quantum state $\ket{\alpha}$:
\begin{equation}
S T^- \ket{\alpha} = P(T^+) S \ket{\alpha} \,,
\label{softSMatrix}
\end{equation} 
where $(T^-, T^+) \in BMS_- \otimes BMS_+$ and $P$ is the antipodal map on the sphere:
\begin{equation}
P(T^+)(\theta, \varphi) = T^+(P(\theta, \varphi)) \,.
\label{conservationLaw}
\end{equation}
 Imposing this invariance implies that if a matrix element $\bra{\beta} S \ket{\alpha}$ is non-vanishing, then
\begin{equation}
	\bra{\alpha} T^- \ket{\alpha} = \bra{\beta}P(T^+) \ket{\beta} \,.
\end{equation}
This means that the memory effect of the outgoing wave, parameterized by $T^+$, must match the memory effect of the incoming wave, parameterized by $T^-$, antipodally at each angle. Because of the constraints \eqref{constraintU} and \eqref{constraintV}, this is equivalent to the statement that the outgoing energy $\Delta \tilde{\mathcal{F}}_{\text{out}}$ matches the ingoing energy $\Delta \tilde{\mathcal{F}}_{\text{in}}$ antipodally at each angle, in particular that $\Delta \tilde{\mathcal{F}}_{\text{out}}$ is fully determined in terms of $\Delta \tilde{\mathcal{F}}_{\text{in}}$.

This criterion has a very interesting connection to IR-physics. As discussed in appendix \ref{app:IR}, we know that in a gapless theory such as gravity most process in which no soft modes are emitted have zero probability \cite{YFS, weinberg}. In order to obtain a finite answer, one has to include the emission of a certain class of soft radiation, namely IR-modes. 
 The sole exception are processes for which the kinematical factor $B_{\alpha, \, \beta}$ defined in \cite{weinberg} is zero. The crucial point is that this happens if and only if the ingoing energy matches the outgoing energy antipodally at each angle, as discussed in detail in \cite{semenoff}. Thus we conclude that\footnote
 {We will elaborate on this point in \cite{usNew}.}
\begin{equation}
		\bra{\alpha} T^- \ket{\alpha} = \bra{\beta}P(T^+) \ket{\beta} \quad \Leftrightarrow \quad  B_{\alpha,\, \beta}=0 \,.
\end{equation}
This means that restricting to processes which fulfill the condition \eqref{conservationLaw} is equivalent to only considering processes that are IR-finite even without including IR-emission.

A priori, there is nothing wrong with solely considering such processes. However, they only form a set of measure zero of those processes that occur in reality. Namely any realistic scattering is accompanied by the emission of soft IR-modes. Once we include soft IR-emission, we know that all processes -- with an arbitrary non-zero value of $B_{\alpha, \, \beta}$ -- are IR-finite. Thus in reality, any process can occur, \ie also ones that do not fulfill the antipodal matching condition \eqref{conservationLaw}. For this reason, we will not restrict ourselves to processes that obey \eqref{conservationLaw}.

\subsubsection*{Role of Soft IR-Gravitons}
 Since we consider processes that include the emission of soft IR-modes, it is natural to ask if those modes could carry information about the black hole state and if they could even suffice to purify black hole evaporation. This is only possible if two conditions are fulfilled. First, IR-modes would have to be sensitive to the microstate of the black hole. We expect this not to be the case since they only depend on the initial and final scattering state, but not on the details of the process.
While we leave the above question for future work, we now focus on the second condition, namely that the number of resolvable IR-modes would have to be big enough to be able to carry the whole black hole entropy.

In contrast to the proposal made in \cite{stromingerNew}, we argue that generic properties of IR-physics imply that this is not the case. As follows from equation \eqref{numberUnresolved},  the number of unresolved soft modes scales logarithmically with the IR-resolution scale. Thus, when we lower the energy scale of resolution from $\epsilon_1$ to $\epsilon_2$, the number of additional IR-modes that we can resolve is:
\begin{equation}
n^{\text{res}}_{\text{soft}} \sim B_{\alpha, \, \beta} \ln \frac{\epsilon_1}{\epsilon_2} \,,
\end{equation}
where $B_{\alpha, \, \beta}\sim G s$ is determined by the energy scale $s$ of the process. 
We apply this formula to the single emission of a Hawking quantum of energy $r_g^{-1}$. It will be crucial in this argument that Hawking radiation gets softer for bigger black holes. The worst resolution scale compatible with observing this process is $\epsilon_1=r_g^{-1}$. The key point is that the resolution scale in this process cannot be arbitrarily good. Namely, it is set by the time-scale of the process, $\epsilon_2\sim t_{\text{b-h}}^{-1}$. Since the life-time of a black hole scales as $t_{\text{b-h}} \sim N r_g $, we get 
\begin{equation}
n^{\text{res}}_{\text{soft}} \lesssim \frac{1}{N}\ln N \,.
\end{equation}
Thus, after the black hole has evaporated by emitting $N$ Hawking quanta, the maximal entropy contained in the soft IR-modes is
\begin{equation}
S_{\text{soft}} \lesssim \ln N \,.
\end{equation}
Independently of the question whether IR-modes are strongly correlated with the Hawking quanta, this shows that they cannot account for the whole entropy of the black hole, but could only give a logarithmic correction. Of course, this leaves open the possibility that non-IR soft modes could account for the bulk of black hole information. However, since they are independent of IR-divergences and accompanying dressing tools, the results of infrared physics do not constrain them.

\subsubsection*{Role of Zero-Energy Gravitons}

 Finally, we briefly discuss the role of zero-energy gravitons. To this end, we consider the process of a Goldstone supertranslation in the limit of zero energy injected and zero energy radiated. This is equivalent to the scattering with a graviton of zero energy.  Since those carry no energy, they cannot emit IR-modes and therefore obey the antipodal matching condition \eqref{conservationLaw}. This fact simply reflects the well-known decoupling of soft modes \cite{porrati1, sever, raoul,mischa, porrati2}. The physical interpretation of this phenomenon is that any bulk configuration is transparent for decoupled soft modes so that the energy profile of the outgoing wave is antipodally related to that of incoming energy. 

But when the emitted/injected radiation does not carry energy, $\mu^\pm=0$, then the constraint \eqref{deltaMU} (or respectively \eqref{deltaMV}) implies that $\int \ldiff[2]{\Omega} \mathcal{F}_{\text{in/out}} =0$. Since $\mathcal{F}_{\text{in/out}}$ represents real gravitational radiation, it follows by the requirement of positive energy that $\mathcal{F}_{\text{in/out}}=0$. Thus, only supertranslations with $D^2(D^2+2) T^\pm = 0$ can occur in such a zero-energy process. This means that only the angular modes $l=0,1$ are left, \ie time- and space-translation. Hence zero-energy radiation cannot lead to a physical memory effect that is observable in finite time. In other words, it is impossible to measure a zero-energy graviton in finite time.

The upshot is that predicting $T^+$ from the knowledge of $T^-$ is only possible for zero-energy modes. Those are, however, unphysical since they cannot be measured in finite time. So we will only consider processes of non-zero energy in our paper. As explained, it is not possible for them to constrain or even predict $T^+$ from $T^-$ without detailed knowledge of the dynamics in the bulk. The response function, which determines $T^+$ in terms of $T^-$, is trivial only for modes of zero energy.

\subsubsection*{Physical Hair With Non-Zero Energy}
\begin{figure}
	\begin{center}
		\includegraphics[width=0.3\textwidth, trim={5cm 6cm 4.6cm 3cm},clip]{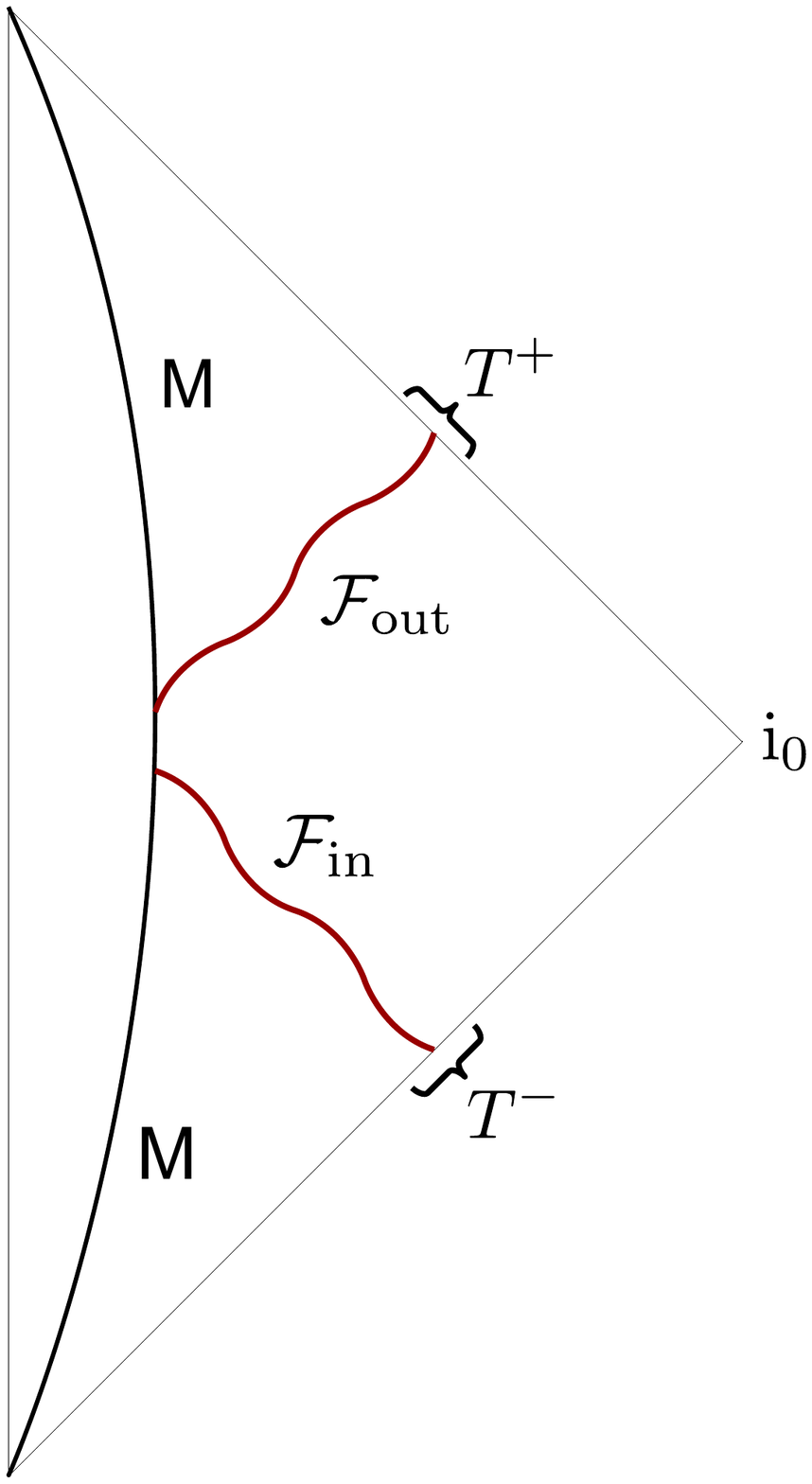}
		\caption{A Goldstone supertranslation on a generic system of mass $M$. Radiation with angular distribution $\mathcal{F}_{\text{in}}$ scatters so that radiation with angular distribution $\mathcal{F}_{\text{out}}$ is returned. Since $\int \ldiff{v} \int \ldiff[2]{\Omega} \mathcal{F}_{\text{in}}=\int \ldiff{u} \ldiff[2]{\Omega} \mathcal{F}_{\text{out}}$, the total energy of the system remains unchanged.  Here $\mathcal{F}_{\text{in}}$ can be described in terms of the supertranslation $T^-$ and $\mathcal{F}_{\text{out}}$ in terms of $T^+$.}
		\label{fig:goldstonePlanet}
	\end{center}
\end{figure}

So from here on, we consider the case where after we inject radiation $\mathcal{F}_{\text{in}}$ of non-zero total energy $\mu$, the system radiates back the same total amount of energy, but with a possibly different distribution $\mathcal{F}_{\text{out}}$.\footnote
{We recall that we restrict ourselves for now to a pure gravitational radiation, which propagates along null geodesics. Therefore, all emitted energy is bound to reach future null infinity $\mathcal{J}^+$.}
While such systems are of course special, we will see that black holes can be one of them. This is a zero-energy process in the sense that the total energy of the system does not change. Thus, this process, which is depicted in figure \ref{fig:goldstonePlanet}, constitutes a transformation between degenerate systems and therefore defines hair.

As far as we reduce ourselves to gravitational radiation, we can generically describe this process in terms of two supertranslations: At $\mathcal{J}^-$, $T^-$ is determined by the angular distribution $\Delta\tilde{\mathcal{F}}_{\text{in}}$ of incoming energy according to the constraint \eqref{constraintV} and at $\mathcal{J}^+$, $T^+$ follows from the angular distribution $\Delta \tilde{\mathcal{F}}_{\text{out}}$ of outgoing energy via the constraint \eqref{constraintU}.  Thus, the whole process is associated to an element $(T^-,\, T^+) \in BMS_{-}\otimes BMS_{+}$. It describes a zero-energy transition which interpolates between two spacetimes of the same total energy, but contrary to the case of a zero-energy mode, this transformation is non-trivial and it is not decoupled. 

 It is crucial to note that for an asymptotic observer, $T^-$ and $T^+$ are independent. Whereas one is free to choose $T^-$ by preparing an appropriate incoming radiation, $T^+$ is sensitive to the properties of the system in the bulk. In other words, $T^+$ is a response of the system which does not only depend on the ingoing radiation, parameterized by $T^-$, but also on the state of the system and its particular dynamics, which are not entirely visible asymptotically. In particular, there is no reason why $(T^-,\, T^+)$ should be in \textit{any} subgroup of $BMS_- \otimes BMS_+$.

\subsubsection*{Coordinate Matching}
\label{ssec:matching}

 In order to compare ingoing and outgoing radiation, \ie $T^-$ and $T^+$,  we need to relate the supertranslation field $C^{-}$ in advanced coordinates to the supertranslation field $C^{+}$ in retarded coordinates.  
Namely, we assume that we are given a classical spacetime whose asymptotic behavior is fully known to us. Then it is possible to describe this spacetime both in advanced and retarded BMS-gauge. Given an advanced coordinate system $g_{\mu\nu}^v$, we want to know if there is a unique retarded coordinate system $g_{\mu\nu}^u$ we can associate to it. If we have such a mapping, it determines the relation of the advanced supertranslation field $C^-$, defined as the $r^1$ part of $g^v_{AB}$, and the retarded supertranslation field $C^+$, defined as the $r^1$ part of $g^u_{AB}$.

Given $g_{\mu\nu}^v$, we therefore have to find a diffeomorphism $\mathcal{D}$ such that $g_{\mu\nu}^u:=\mathcal{D}(g_{\mu\nu}^v)$ is in retarded BMS-gauge. Then we can read off from $g_{\mu\nu}^u$ the $C^+$ associated to $C^-$. However, we could have instead considered the diffeomorphism $\mathcal{D}'= T^+\circ\mathcal{D}$, where $T^+$ is a supertranslation diffeomorphism in retarded coordinates. Also $\mathcal{D}'$ transforms the metric in advanced BMS-gauge to a metric in retarded BMS-coordinates. Clearly, if $T^+$ is a nontrivial supertranslation, the supertranslation field in the resulting metric differs from the one in $g^u$. From this consideration it is obvious that the matching between the advanced and the retarded supertranslation field is in general not unique. 

The only hope we could have is that there is a natural way to identify $C^-$ and $C^+$.  In a static situation, a natural prescription is to require that the spatial part of the two metrics matches, \ie
\begin{equation}
g^u_{AB} = g^v_{AB} \,.
\label{matchingPrescription}
\end{equation}
As is shown explicitly in appendix \ref{app:matching} for the example of the Schwarzschild metric, we can achieve this by identifying $C^+(\theta, \varphi) = - C^-(\theta, \varphi)$, as also proposed in \cite{compereMatching}. Up to a sign, we match the supertranslation field angle-wise.   Consequently, the same matching holds for the supertranslations:
\begin{equation}
T^+(\theta, \varphi) = - T^-(\theta, \varphi) \,.
	\label{matching}
\end{equation}

There are several reason why the coordinate matching \eqref{matching} is natural. First of all, the prescription \eqref{matchingPrescription} comes from a simple intuition. For an observer in a static spacetime who lives on a sphere of fixed radius, the description of the sphere should be the same independently of the choice of time coordinate. More generically, it is possible to require that the action of advanced and retarded supertranslations is the same in the bulk. This was done in \cite{compereFinite, compereMatching} for the cases of Schwarzschild and Minkowski.

Moreover, we can consider a detector at big radius which is sensitive to gravitational memory. Then we investigate a process of back scattering, in which the angular distributions of incoming and outgoing energy are identical at each angle. This corresponds to a wall in the bulk which reflects the wave without further modifying it. In this case, the memory effect the ingoing wave causes, parameterized by $T^-$, is exactly canceled by the memory effect of the outgoing wave, parameterized by $T^+$, so that there is no overall memory effect after the process. In that case, if we match $T^-$ and $T^+$ at each angle as in \eqref{matching}, it is possible to simply describe the overall memory effect as $T^-+T^+$.

However, it is crucial to stress that the coordinate matching \eqref{matching} does not have any constraining power on the physical process. It does not predict outgoing from ingoing radiation, but only shows how one and the same setup can be described in different coordinates. This is also evident from figure \ref{fig:goldstonePlanet}. The matching condition at $i^0$ only relates the absolute values of the supertranslation fields. In contrast, processes of non-zero energy solely determine a change of the supertranslation field, as is clear from equations \eqref{constraintU} and \eqref{constraintV}. Thus, radiation of non-zero energy is independent of the coordinate matching.

\section{Application of Hair}
\label{sec:implanting}
\subsection{Planetary Hair}

In order to make the ideas presented above concrete, we discuss an explicit example, namely the  application of a Goldstone supertranslation to a certain class of planets. We start from a spherically symmetric nongravitational source $T_{\mu\nu}$, which sources a spherically symmetric spacetime $g_{\mu\nu}$ with ADM-mass $M$. In such a spacetime, we want to realize a Goldstone supertranslation, \ie we send in a wave with total energy $\mu$ and angular distribution $\Delta\tilde{\mathcal{F}}_{\text{in}}$ in such a way that after some time, the planet emits a wave of the same energy $\mu$ but with a possibly different angular distribution $\Delta\tilde{\mathcal{F}}_{\text{out}}$. Of course, only a special class of planets behaves in that way.

We explicitly construct such spacetimes in appendix \ref{app:planetSolution}, to which we refer the reader for details of the calculation. First, we consider the incoming wave. As discussed, the angular distribution $\Delta\tilde{\mathcal{F}}_{\text{in}}$ of injected energy determines an advanced supertranslation $T^-$. As derived in equation \eqref{dynamicalTransformationV}, we can use it to describe the change of the metric due to the injected radiation:
\begin{align}
\delta g^v_{\mu\nu} = \tau_{v_0,\,v_1}(v) s^-(r) \left(\mathcal{L}_{\xi_v(T^-)} g^v_{\mu\nu} + \frac{2 \mu G}{r} \delta_\mu^0 \delta_\nu^0 \right) \,, \label{planetInjected}
\end{align}
where $\mathcal{L}_{\xi_v(T^-)} g^v_{\mu\nu}$ is an infinitesimal supertranslation which changes the supertranslation field by a small amount $T^-$. Whereas the asymptotic supertranslation $T^-$ only depends on the leading part of the incoming energy, it is crucial to note that the transformation \eqref{planetInjected} also depends on a careful choice of the subleading components of the incoming wave.\footnote
{Subleading terms are the $1/r^3$-term in $T_{00}$ and the whole $T_{0A}$ in \eqref{waveSchwarzschild}. If one does not insist that the wave acts as a supertranslation also in the bulk, one is free to choose the coefficient of one of the two terms. The other one is determined by energy conservation: $T_{\mu\nu}^{\phantom{\mu\nu};\mu}=0$.}
Only with a particular choice, the wave acts as a diffeomorphism not only asymptotically but also in the bulk outside the planet.

We observe that the effect of the wave is twofold. First, it adds the total mass $\mu$ to the planet and secondly, it supertranslates the metric by $T^-$. However, these effects are localized both in space and time. The function $\tau_{v_0,\,v_1}(v)$ describes the smooth interpolation between $g^v_{\mu\nu}$ and $g^v_{\mu\nu} + \delta g^v_{\mu\nu}$, \ie we have $\tau_{v_0,\,v_1}(v<v_0)=0$ and $\tau_{v_0,\,v_1}(v>v_1)=1$. The function $s^-(r)$ describes the absorption of the wave, namely absorption takes place whenever $s^{-\,'}(r)<0$. There is no absorption outside the planet, \ie $s^-(r>R)=1$, where $R$ is the radius of the planet, and the wave is fully absorbed before it reaches the center, $s^-(r=0)=0$. It will be crucial to note that the transformation $s^-(r)\mathcal{L}_{\xi_v(T^-)} g^v_{\mu\nu}$ only acts as a diffeomorphism when $s^{-\,'}(r)=0$. 

Moreover, the transformation \eqref{planetInjected} shows that we focus on planets which have a second very special property aside from the fact that they emit as much energy as they receive: Namely there is no transport of energy between different angles. This means that the mass of the planet does not redistribute after absorption (the same will be true after emission).  The fact that this assumption is unnatural and not true for generic systems will contribute to our conclusions.

As a second step, we consider the emission of a wave by the planet.  Of course, the properties of the emitted wave depend on the internal dynamics of the source $T_{\mu\nu}$. It is crucial to note we cannot resolve them in our purely gravitational treatment, \ie we cannot predict what wave will be emitted. From the point of view of gravity, any emission process is possible as long as it respects energy-momentum-conservation. However, we can study the effect of a given emitted wave. As derived in equation \eqref{dynamicalTransformation2}, it can be described in terms of the supertranslation $T^+$ induced by the angular distribution $\Delta\tilde{\mathcal{F}}_{\text{out}}$ of outgoing energy:
\begin{align}
\delta g^u_{\mu\nu} = \tau_{u_0,\,u_1}(u)s^+(r) \left(\mathcal{L}_{\xi_u(T^+)} g^{u}_{\mu\nu} - \frac{2\mu G}{r} \delta_\mu^0 \delta_\nu^0 \right) \,.
\label{planetEmitted}
\end{align}
As for the case of absorption, the emission has two effects: It decreases the total mass by $\mu$ and it supertranslates the metric by $T^+$. Moreover, it is localized in space and time in an analogous manner. 

We want to compare the planet before and after the Goldstone supertranslation, \ie we are interested in the combined effect of the transformations \eqref{planetInjected} and \eqref{planetEmitted}. To this end, we have to specify a mapping between the advanced and retarded supertranslations. As explained in section \ref{ssec:matching}, we employ the angle-wise matching \eqref{matching}. Thus, we obtain the static final state of the planet:
\begin{align}
\delta g^{\text{tot}}_{\mu\nu} & =  \theta(r-R) \mathcal{L}_{\xi_u(T^+-T^-)} g_{\mu\nu} \nonumber\\
& + \theta(R-r)\left(s^+(r)\mathcal{L}_{\xi_u(T^+)}g_{\mu\nu} - s^-(r) \mathcal{L}_{\xi_u(T^-)} g_{\mu\nu}  \right)\,.
\label{planetTransformation}
\end{align}
 We get a planet which has the same ADM-mass but a different angular distribution of mass. This is clear from the fact that the transformation \eqref{planetTransformation} acts as a diffeomorphism only outside the planet. 
 
 Since we used in our computation a planet with the special property that its angular distribution of energy is frozen, we can read off the distribution from difference of energy distributions of the injected and emitted wave. In this case, $T^- - T^+$ encodes all information about the angular energy distribution of the planet in the bulk.\footnote
{For the planet with frozen energy distribution, there is also a very literal way in which one can interpret the quantity $T^- - T^+$: One can imagine a gedankenexperiment where a source of light is located in the interior of the planet after the Goldstone supertranslation and we collect the light rays on the sky. The light sent from this common center point determines in this way a section at infinity described by the supertranslation field $T^- - T^+$.  Thus, the different redshift effects due to the inhomogeneities of the planet matter distribution define a supertranslated section in the sky as the one for which light rays originate from a common spacetime point. This is reminiscent of Penrose's concept of "good sections" \cite{penrose}.}
However, this is no longer true for generic systems which exhibit non-trivial dynamics after absorption and emission. In that case, $T^-$ and $T^+$ merely encode the initial state. Only with full knowledge of the theory which governs the internal dynamics of the planet, we can infer the state of the planet at a later time from the asymptotic data $T^-$ and $T^+$. 

\subsubsection*{The Role of Supertranslations}
 In summary, we obtain the following key properties of a Goldstone supertranslation in the case of a planet: Outside the planet, it acts as a diffeomorphism. In particular, it does not change its ADM-mass. In contrast, it does not act as a diffeomorphism inside the planet where absorption takes place. Therefore, it is not a trivial global diffeomorphism but changes the spacetime physically. Thus, the Goldstone supertranslation encodes differences in the angular distribution among matter configurations degenerate with respect to the ADM-conserved quantities. 

It is crucial to discuss the role of supertranslations in this process:
\begin{itemize}
	\item For an asymptotic observer, $(T^-,\,T^+)$ can be used as label for the angular features of ingoing and outgoing radiation. 
	\item An asymptotic observer, however, cannot infer $T^+$ from $T^-$. This is only possible with knowledge of internal dynamics of the planet.
	\item Thus, $(T^-,\,T^+)$ is a bookkeeping tool but without detailed information about the interior, it does not have predictive power.
\end{itemize}
As we shall discuss in a moment, the same conclusions hold in the black hole case. The only difference is that the internal dynamics leading to emission are fully quantum mechanical for a black hole. This will mean that in any classical description, supertranslation cannot constrain or even predict black hole evaporation.

 Using the example of the planet, it is easy to convince ourselves that antipodal matching cannot play a role in processes of non-zero energy. Namely if it did, this would mean that the only planets which could exist would have the extremely special property that they emit all energy they receive from one side exactly on the other side.

\subsubsection*{Hidden Angular Features}
Finally, we discuss the transformation \eqref{planetTransformation} when we do not have access to $(T^-,\,T^+)$, \ie when we do not record ingoing and outgoing radiation but only compare the initial and final state of the planet. In that case, the planet possesses an interesting property, namely a special kind of no-hair-theorem. Concretely, we take the perspective of an observer who has no access to the interior of the planet and discuss the difference between two planets which have the same mass but a different angular mass distribution. As we have observed, the transformation \eqref{planetTransformation} acts as a diffeomorphism outside the planet. Therefore, an outside observer cannot distinguish the two following cases when he is given a supertranslated outside metric. First, it could be the result of the transformation \eqref{planetTransformation}, where the planet was physically changed due to a Goldstone supertranslation. Secondly, however, one can also obtain the supertranslated metric by acting on the initial planet with a global diffeomorphism. In this case, clearly, the planet does not change. Thus, also for a planet, an outside observer is not able to resolve angular features. In order to decide whether two asymptotic metrics differing by a supertranslation describe two different distributions of matter or the same distribution of matter in different coordinates, one needs access to the whole spacetime, \ie the interior of the source.

We conclude that generic gravitational systems posses physical angular features which are inaccessible for an outside observer. This strengthens our believe that the microstates of a black hole have a non-trivial projection on angular features. The only difference is that while the restriction to outside measurements was artificial in the case of the planet, an outside observer has {\it in principle} no access to the interior of a black hole. As we will discuss in the next section, he can therefore never decide whether a supertranslated metric corresponds to a physical change of the matter inside the black hole or to a global and therefore meaningless diffeomorphism. This is the reason for the classical no-hair theorem of a black hole and why we assign an entropy to the black hole and not to the planet.

\subsection{Black Hole Quantum Hair}

\subsubsection*{Supertranslations as Bookkeeping Device}
Now we are ready to discuss the system of our interest, namely black holes. Since absorption and emission are of different nature in that case, we will discuss them separately. For absorption, we can proceed in full analogy to the planet and inject a wave with total energy $\mu$ and arbitrary angular distribution $\Delta\tilde{\mathcal{F}}_{\text{in}}$. By Birkhoff's theorem, the spacetime outside the black hole is the same as for the planet so that the wave behaves identically. As in the case of the planet, the wave cannot be absorbed outside the horizon  and acts as a diffeomorphism everywhere outside the black hole and also on the horizon.

 For the planet, we observed that the knowledge of injected energy alone does not suffice to predict what radiation the planet emits. Instead, this can only be done with knowledge of the interior dynamics of the planet. Those, however, can be described classically in the case of the planet. For the black hole, the situation is even worse. Not only do we not have access to any interior dynamics, but these dynamics are also fully quantum. It is impossible to describe them even with full classical knowledge of the interior of the black hole. 
	
 Before we elaborate on this point, we first show how it is possible to use supertranslations as bookkeeping device for black hole evaporation. Unlike for the case of the planet, this is a non-trivial question since the evaporation products are generic quantum states.  In order to define an associated supertranslation, we shall proceed as follows. We consider an ensemble of {\it quantum-mechanically} identical black holes of mass $M$.\footnote
 {Experimentally, we can realize this by preparing identical quantum states in such a way that they collapse and form black holes.}
For each black hole, we wait until it has emitted exactly one Hawking quantum.  We only record their angular features, \ie the deviation from an isotropic emission. This means that we assume that the microstates of the black hole have a non-trivial projection on angular features of the evaporation products. As explained in the introduction and illustrated for the case of the planet, we believe this assumption to be natural. Thus, we record the Hawking quanta using a filter for angular features, where we use one for each spherical mode $(l,m)$. This defines a probability distribution for the angular features of the ensemble:
 \begin{equation}
 P(l,m) \,.
 \label{probabilities}
 \end{equation}
 Obviously, the probability distribution \eqref{probabilities} only contains a part of the quantum-mechanically available information. However, we will only focus on it since it can be described in terms of classical supertranslations. At this point, it is crucial to point out that the probability distribution \eqref{probabilities} does not originate from a mixed state but as a result of an ordinary quantum measurement. Thus, unlike in a description in terms of a mixed state, it is not associated to any fundamental loss of information.
 
 Since we need to recover a featureless emission in the semi-classical limit, it follows that 
 \begin{equation}
 P(0,0)=1-\epsilon\,,
 \label{probabilityDeviation}
 \end{equation} 
 where $\epsilon \rightarrow 0$ in the semi-classical limit. This means that only a fraction $\epsilon$ of the emitted quanta carries features. For $l\geq 2$, we consequently get
 \begin{equation}
 	P(l,m) = \epsilon A_{l,m} \,,
 	\label{probabilityFeature}
 \end{equation}
 where $\sum_{l=2}^\infty \sum_{m=-l}^{m=l} A_{l,m} =1$. The information contained in the $P(l,m)$ is purely quantum mechanical. At the semi-classical level, we have that $P(l,m)=\delta_{l0}$ and in the classical limit, we have no emission at all. 
 
\begin{figure}
	\begin{center}
		\includegraphics[width=0.3\textwidth, trim={5cm 6cm 5cm 8.6cm},clip]{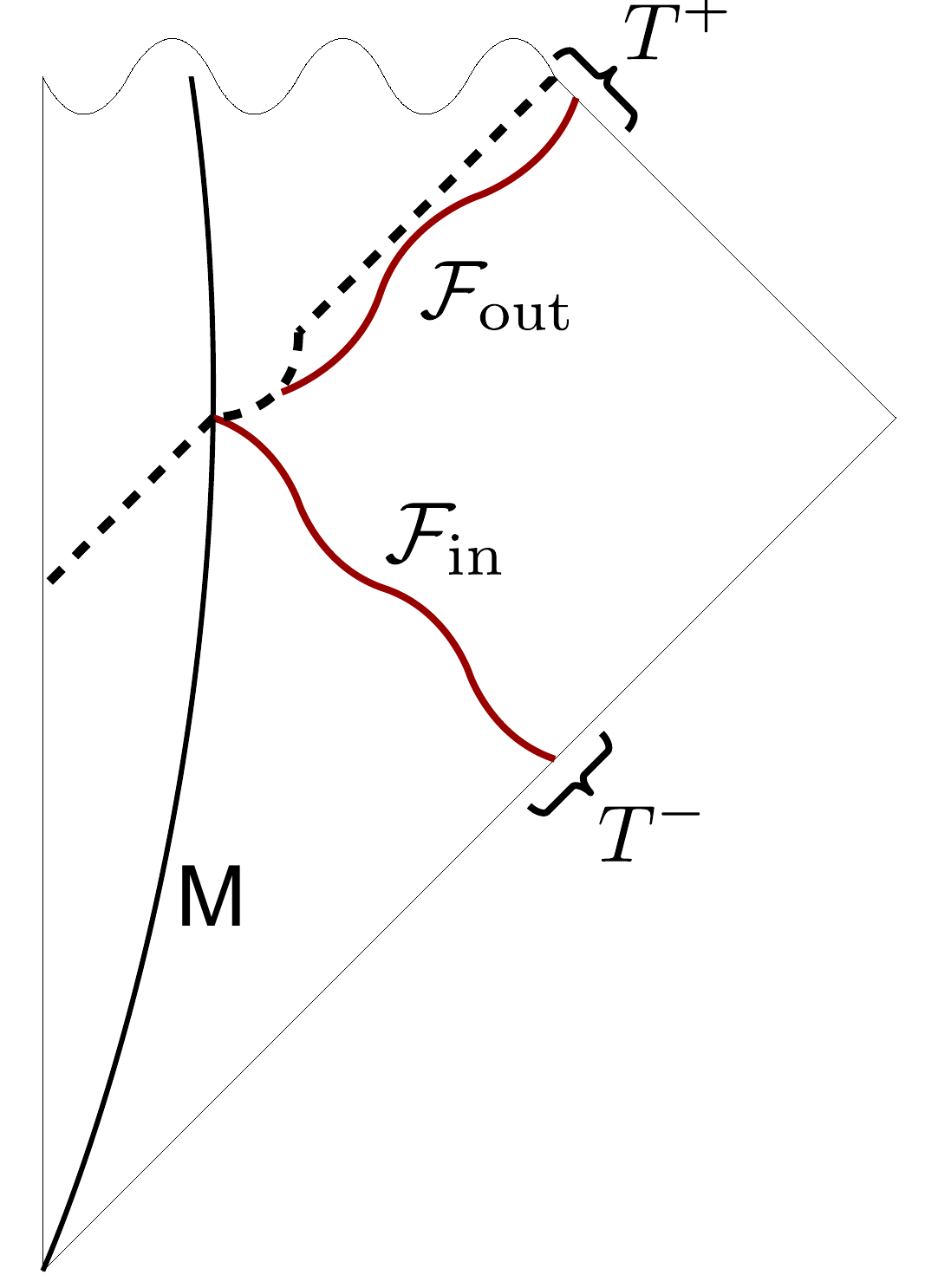}
		\caption{A Goldstone supertranslation on a black hole of mass $M$. First, it absorbs radiation with angular distribution $\mathcal{F}_{\text{in}}$ and then it evaporates radiation with angular distribution $\mathcal{F}_{\text{out}}$. Since $\int \ldiff{v} \int \ldiff[2]{\Omega} \mathcal{F}_{\text{in}}=\int \ldiff{u} \ldiff[2]{\Omega} \mathcal{F}_{\text{out}}$, the total energy of the black hole remains unchanged. Here $\mathcal{F}_{\text{in}}$ can be described in terms of the supertranslation $T^-$ and $\mathcal{F}_{\text{out}}$ in terms of $T^+$.}
		\label{fig:goldstoneBlackHole}
	\end{center}
\end{figure}

 Using the quantum probability distribution \eqref{probabilities}, we can associate to every Hawking quantum an average energy flux:
 \begin{equation}
 \mathcal{F}_{\text{out}} = \hbar r_g^{-1} \sum_{l=0}^\infty \sum_{m=-l}^{m=l}  P(l,m) Y_{l,m} \,,
 \label{quantumFlux}
 \end{equation}
 where $Y_{l,m}$ are the standard spherical harmonics. Just like for the case of the planet, where we considered a classical process of emission, we can use the flux \eqref{quantumFlux} to define a {\it classical} supertranslation $T^+$. Of course, this is only possible as long as $\hbar \neq 0$ since the energy flux is zero otherwise.
 When we record the quantum-mechanically emitted energy $\mathcal{F}_{\text{out}}$, we can proceed in analogy to the planet and use the supertranslation fields $T^-$ and $T^+$ to track the evolution of the black hole. Concretely, in order to perform a Goldstone supertranslation, we first inject an energy $\mu$ and then we wait until $n_H=\mu/(\hbar r_g^{-1})$ quanta have evaporated, as is depicted in figure \ref{fig:goldstoneBlackHole}. Then we end up with a black hole of the same mass as before the process. Of course, the sensitivity of the final state on the initial state is suppressed by $\mu/M$ but unitarity dictates that the dependence is never trivial.

\subsubsection*{Insufficiency of Supertranslation Hair}
 However, it is impossible to predict $T^+$ solely from the knowledge of $T^-$. The reason is that the wave that we inject acts as a diffeomorphism outside the horizon and also on the horizon. Therefore, the geometry outside the black hole is unchanged after the wave has passed. Since the semi-classical Hawking calculation is only sensitive to the geometry on the horizon and outside the black hole, its result cannot change as a result of a supertranslation diffeomorphism. Therefore, additional knowledge about the interior is required to predict $T^+$.
 
 We can make this argument more concrete by taking the perspective of an observer who lives in a Schwarzschild metric supertranslated by $T^-$. The observer has no record of how the black hole was formed and is only allowed to make experiment outside the horizon. Her goal is to determine the microstate of the black hole. More specifically, she wants to know if the black hole is in the bald microstate, whose evaporation products are featureless and in particular perfectly isotropic, or in a non-trivial microstate, whose evaporation products carry some angular features. By our definition of microstate, one way to do so is to wait till the black hole has evaporated and to determine the properties of the evaporation products. 
 
 The question we are asking is if there is another way to determine the microstate of a black hole. The answer is negative, for the following reason: When an outside observer finds herself in a black hole metric with supertranslation field $T^-$, this can happen because of two very distinct reason. Firstly, it could be the result of injecting a wave with a non-trivial angular distribution of energy into a black hole. In that case, the black hole is in a non-trivial microstate and $T^-$ indeed characterizes the microstate. 
 
 However, there is a second way in which we can obtain a supertranslated Schwarzschild metric. Namely, we can consider a featureless microstate, whose evaporation products are isotropic, and apply a supertranslation diffeomorphism to this setup. In this way, we do not change the physical state of the black hole but only describe it in a different metric. Thus, $T^-$ can also correspond to a featureless microstate described in different coordinates.
 
 Without access to the evaporation products, the only way to distinguish those two cases -- injection of wave with angular features versus global diffeomorphism -- is to enter the black hole. There, the wave acts non-trivially, \ie not as a diffeomorphism, whereas the global diffeomorphism still does.
 Since the same exterior metric can correspond to both a trivial and a non-trivial microstate, the metric alone cannot suffice to predict the evaporation products.  From the outside, it is therefore impossible to distinguish classical supertranslation hair and global diffeomorphisms.

 In summary, as in the case of a planet, we can use $(T^-,\,T^+)$ as a natural bookkeeping device for the black hole to track the angular features of ingoing and outgoing radiation. However, knowing $T^-$ does not suffice to predict $T^+$, \ie an observer outside the black hole cannot infer $T^+$ from $T^-$. This is only possible with a microscopic model of the interior dynamics of the black hole.

\subsubsection*{Generalization to Evaporation}
Having discussed how we can implant hair on a black hole with a Goldstone supertranslation, it is trivially to consider the case of pure evaporation. We obtain it if we just leave out the first part of the Goldstone supertranslation, namely the injection of a wave. Therefore, it suffices to consider $\mathcal{J}^+$ as screen, where the constraint \eqref{constraintU} determines the retarded supertranslation field $T^+$ in terms of the angular distribution $\Delta\tilde{\mathcal{F}}_\text{out}$. In that case, the metric outside the black hole changes according to \eqref{planetEmitted}:
\begin{align}
\delta g^u_{\mu\nu} = \tau_{u_0,\,u_1}(u) \left(\mathcal{L}_{\xi_u(T^+)} g^{u}_{\mu\nu} - \frac{2\mu G}{r} \delta_\mu^0 \delta_\nu^0 \right) \,.
\end{align}
This equation shows that the back reaction splits in two parts. First, energy conservation dictates that the mass of the black hole is reduced by the total emitted energy $\mu = \int \ldiff{u} \int \ldiff[2]{\Omega}  \mathcal{F}_{\text{out}}$. This part of the back reaction is undebatable but does not suffice to ensure unitarity of the process. Fortunately, $\mathcal{F}_{\text{out}}$ contains more information than just the emitted energy, namely the supertranslation $T^+$. Consequently, we obtain the back reacted black hole not only by reducing its mass, but by supertranslating it by $T^+$.  This approach is only valid if the supertranslation acts non-trivially in the interior of the black hole, \eg because it is induced by a physical wave. But in that case, the ability to associate hair to a black hole is equivalent to the ability to purify its evaporation.

\subsection{A Comment on Page's Time}

So far, we have not specified the magnitude of deviations from a thermal evaporation. We can estimate them by requiring that we reproduce Page's time in our approach. In its most basic formulation, Page's time is a direct consequence of describing the black hole evaporation in a Hilbert space of fixed dimension. In brief, if we keep the dimension of the full Hilbert space, which describes at any time both the black hole and the emitted radiation, fixed and equal to $2^N$, then at $t=t_P$, which corresponds {to the half life-time, \ie the evaporation of $\sim N/2$ quanta}, there is no place to continue increasing the entanglement between the radiation and the black hole internal degrees of freedom. At this time, entanglement starts to decrease and information starts to be delivered. This makes clear why purification of black hole evaporation relies on fixing $N$ and keeping it finite.

 Page's time can be defined as the time-scale for the emission of the order of $N$ quanta. Therefore, we first consider an ensemble of $N$ identical quantum mechanical black holes and for each of them, we record the first emitted quantum.
For a measurement on a single black hole, the standard deviation is
\begin{equation}
\sigma_1 \sim O(1) 
\end{equation}
since the quanta are distributed isotropically to leading order. However, when we average over $N$ measurements, the standard deviation decreases as
\begin{equation}
\sigma_N\sim \frac{1}{\sqrt{N}}\,.
\end{equation}
Features become visible as soon as their strength becomes bigger or equal than the uncertainty of the measurement. After Page's time we can therefore resolve features with the relative amplitude
	\begin{equation}
	\epsilon \sim	\frac{1}{\sqrt{N}} \,.
	\end{equation}
	In the formulation of the probability distribution \eqref{probabilityDeviation}, this means that after $O(N)$ measurement, those features becomes visible which are only carried by a fraction $1/\sqrt{N}$ of the quanta. 

 So far, we have only considered one emission for $N$ identical black holes. If we consider instead $O(N)$ emissions of a single black hole, the difference is that the probability distribution for each emission step is generically different. This is true because of the back reaction of the previously emitted quanta. However, the argument in terms of the resolution stays the same, \ie after Page's time, we can still resolve those features which are only carried by a fraction $\epsilon \sim 1/\sqrt{N}$ of quanta. This argument provides evidence for the black hole $N$-portrait \cite{NPortrait} where features are $1/N$-effects with a resolution scale $O(1/{\sqrt{N}})$.\footnote
{An interesting question that we shall not discuss in this note but that can be worth to mention is the possibility that a quantum computer designed using a Grover like algorithm \cite{grover} can reduce $t_P$ from $O(N)$ to $O(\sqrt{N})$.}
 In particular, in an $S$-matrix analysis along the lines of \cite{2toN}, where black hole formation was studied as $2\rightarrow N$-scattering process, angular features should appear as $1/N$-correction to the leading amplitude.

\section{Conclusion}
\label{sec:conclusion}

The main message of this note is easy to summarize. Any form of black hole hair should imply the existence of features in the black hole evaporation products, \ie in the emitted radiation. This obvious requirement immediately entails, given the intrinsically quantum nature of black hole radiation, that black hole hair should be defined quantum-mechanically and that such a definition is inseparable from the mechanism through which the black hole delivers, in the radiation, information about its internal structure.

  In this note, we have suggested to define hair on the basis of elementary processes of classical absorption followed by quantum emission. Moreover, we specialized to angular features in the radiation. This simplification has been done in order to use the asymptotic symmetry group and the corresponding supertranslations to parametrize both the incoming and the emitted radiation. Since we have complete control over the angular features of the injected radiation, we can define hair on the basis of the angular features of the quantum-mechanically emitted radiation. These features encode information about the internal structure of the black hole which can be measured by an external observer. In this sense, they provide an operative and intrinsically quantum definition of hair. 
  
  In principle, we can imagine two different sources of those features of the emitted radiation. The first one is a classical modification of the near horizon geometry that will modify the corresponding semi-classical Bogoliubov transformations. The second one is a real quantum interaction of the injected radiation with the quantum constituents of the black hole. The first possibility requires to define local changes of the horizon geometry that preserve all the ADM-charges. Thus, locally, they can be always tuned to be equivalent to a diffeomorphism. Therefore, they  cannot have observable consequences, \ie classical supertranslations do not suffice to define observable black hole hair. So the only real possibility of quantum emitted radiation with features is having a non-trivial scattering between the injected radiation and the microscopic constituents of the black hole. This means that the features that define hair in the way we are suggesting depend on the microscopic quantum structure, which we can parametrize as a dependence on the black hole entropy $N$. Thus, the hair that we are defining vanishes in the limit $N=\infty$. 
  
  As it is clear from the discussion, this way of addressing the definition of hair is what we can call an $S$-matrix approach, where by that $S$-matrix we simply mean the dynamics involved in the complex process of actual absorption and quantum emission. If we focus on angular features, we can encode the properties of the hair in terms of the commutators, as operators, of this $S$-matrix and the generators of the asymptotic symmetry group. Associating with the injected energy a supertranslation $T^-$ in $BMS_{-}$, a way to approach the existence of hair is by considering the commutator $[S, T^-]$. Generically, the non-trivial hair will be associated with the symmetry generators that are broken since those are the ones that will create net differences between the angular features of the injected and emitted radiation. Although the infrared dynamics of gravity selects the zero-energy modes as natural symmetries of $S$, they are not able to tell us anything about the internal structure of the black hole since they are decoupled. Zero-modes are unable to encode observable features.
  
  What we have presented in this note is just the general framework to address the problem of quantum hair. In order to go further, it is necessary to use a concrete model of the black hole interior. The model in \cite{NPortrait} provides, in principle, the tools to  address this questions in a quantitative way, something to which we hope to come back in the future.

\appendix
\section{Recap of IR-Effects}
\label{app:IR}

In this appendix, we shall collect some well-known facts about infrared physics which could be useful to clarify some controversial aspects on the meaning of soft modes. Some of these issues have been revisited recently in a series of papers \cite{ stromingerMasslessQED, mohd, campiglia, stromingerQED, porrati1, sever, stromingerLecture, raoul, stromingerRevisited}. 
\begin{itemize}	
	\item In QED, asymptotic physical states associated with freely moving charged particles should be dressed in order to satisfy the Gauss law constraints. This dressing simply adds to the freely moving charge its companion electrostatic field, \ie the non radiative part of the retarded Lienard-Wiechert-field behaving at large distances as $1/r^2$. In quantum field theory, this dressing can be defined using a coherent state of off-shell photons \cite{raoul} with dispersion relation $\omega(k) = \vec{k}\vec{v}$ for $\vec{v}$ the velocity of the charged particle. This coherent state dressing contains an infinite number of $k=0$ modes and it is identical to the dressing operator defined in \cite{FK}. In scattering theory, one can define physical asymptotic states and an IR-safe $S$ matrix using this dressing operator.
	\item  Alternatively, one can use no dressing. Then, in perturbative QED as well as in perturbative gravity, we find IR-divergences due to virtual photon/graviton loops. These, after a careful analysis of overlapping divergences, can be resummed and exponentiated \cite{YFS, weinberg}.  When we consider the transition from an initial state $\ket{\alpha}$ to a final state $\ket{\beta}$, we obtain
\begin{equation}
S_{\alpha,\, \beta}^{\text{loop}} = \ex^{B_{\alpha,\, \beta} \ln \frac{\lambda}{\Lambda}/2} S_{\alpha,\, \beta}^0 \,,
\label{softLoop} 
\end{equation}
where $S_{\alpha,\, \beta}^0$ is the amplitude without taking into account soft loops whereas $S_{\alpha,\, \beta}^{\text{loop}}$ contains them. Here $\Lambda$ is a UV-cutoff that defines what is soft, $\lambda$ is a IR-cutoff and $B_{\alpha,\, \beta}$ is a non-negative number, which only depends on the initial state $\ket{\alpha}$ and the final state $\ket{\beta}$. It is zero if and only if the ingoing current in $\ket{\alpha}$ matches the outgoing current in $\ket{\beta}$ antipodally at each angle. In the case of gravity, it scales as $B_{\alpha,\, \beta}\sim G s$, where $s$ is the energy of the process.\footnote
{That this scaling also holds for graviton scattering at an ultra-Planckian center of mass energy was shown in \cite{veneziano}.}
  For $B_{\alpha, \, \beta} \neq 0$, soft loops clearly lead to a vanishing amplitude in the limit $\lambda \rightarrow 0$.
\item In order to cancel the IR-divergences due to virtual photons/gravitons, the Bloch-Nordsieck-recipe \cite{BN} requires to add a certain class of soft emission processes. Again the effects of emitting theses soft IR-modes of energies below $\epsilon$ can be resummed and exponentiated, yielding the rate \cite{YFS, weinberg}
 \begin{equation}
 |S_{\alpha,\, \beta}^\text{full}|^2 := \sum_{\gamma}  |S_{\alpha,\, \beta\gamma}^\text{soft}|^2 =  \ex^{B_{\alpha,\, \beta}\ln\frac{\epsilon}{\lambda}} f(B_{\alpha,\, \beta})  |S_{\alpha,\, \beta}^\text{soft}|^2 \,,
 \label{finiteRate}
 \end{equation}
 where $f(B_{\alpha,\, \beta})$ is due to energy conservation and $f(B_{\alpha,\, \beta})\approx 1$ for small $B_{\alpha,\, \beta}$. Combing the contribution from \eqref{softLoop} and \eqref{finiteRate}, one obtains a rate which is independent of the IR-cutoff $\lambda$ and in particular finite for $\lambda \rightarrow 0$. 
  This cancellation leads to the connection, highlighted in \cite{FK}, between the soft photon theorem and the electrostatic coherent state dressing. In QED, we do not have new symmetries besides the decoupling of zero-energy photons. The same is true in perturbative gravity.	
\item In the correction factor $\ex^{B_{\alpha,\, \beta}\ln\frac{\epsilon}{\lambda}}$ in \eqref{finiteRate}, the $n^{th}$ summand of the exponential series comes from the emission of $n$ IR-modes. Therefore, we can estimate the number of soft modes from the term which gives the biggest contribution in the series. This gives
\begin{equation}
	n_{\text{soft}}^{\text{unres}} \sim B_{\alpha,\, \beta} \ln\frac{\epsilon}{\lambda} \,.
	\label{numberUnresolved}
\end{equation}
We conclude that the number of unresolved soft modes only scales logarithmically with the
infrared resolution scale $\epsilon$.
\end{itemize}

\section{Matching in Schwarzschild Coordinates}
\label{app:matching}
In this section, we demonstrate explicitly how we can transform a Schwarzschild metric with non-trivial supertranslation field from advanced to retarded coordinates. In this way, we show how we can naturally identify the advanced supertranslation field $C^-$ with the retarded one $C^+$. We start from the Schwarzschild metric $g_{\mu\nu}^{v\,,0}$ in advanced coordinates without supertranslation field:
\begin{align}
\mathrm{d} s^2 = -(1-\frac{2GM}{r})\diff v^2 + 2 \diff v\diff r +r^2 \diff \Omega^2 \,.
\end{align} 
The corresponding generators of supertranslations are
\begin{subequations} \label{advancedSupertranslationSchwarzschild}
	\begin{align}
	\xi_v^v =& f^- \,,  \\ 
	\xi_v^r =& -\frac{1}{2} D^2 f^- \,, \\
	\xi_v^A =& \frac{f^{-,\,A}}{r}  \,,
	\end{align}
\end{subequations}
which are characterized by an arbitrary function $f^-$ on the sphere. Thus, the supertranslated metric is
\begin{equation}
	g_{\mu\nu}^v(f^-) = g_{\mu\nu}^{v\,,0} + \mathcal{L}_{\xi_v(f^-)} g_{\mu\nu}^{v\,,0} \,.
	\label{schwarzschildAdvanced}
\end{equation}

In retarded coordinates, the Schwarzschild metric $g_{\mu\nu}^{u\,,0}$ without supertranslation field is:
\begin{align}
\mathrm{d} s^2 = -(1-\frac{2GM}{r})\diff u^2 - 2 \diff u\diff r +r^2 \diff \Omega^2 \,.
\end{align} 
The corresponding generators of supertranslations are
\begin{subequations}
	\begin{align}
	\xi_u^v =& f^+ \,,  \\ 
	\xi_u^r =& \frac{1}{2} D^2 f^+ \,, \\
	\xi_u^A =& -\frac{f^{+,\,A}}{r}  \,,
	\end{align}
\end{subequations}
where it is important to note that the signs of $\xi_u^r$ and $\xi_u^A$ have changed with respect to \eqref{advancedSupertranslationSchwarzschild}. The supertranslated metric is:
\begin{equation}
g_{\mu\nu}^u(f^+) = g_{\mu\nu}^{u\,,0} + \mathcal{L}_{\xi_u(f^+)} g_{\mu\nu}^{u\,,0} \,.
\label{schwarzschildRetarded}
\end{equation}

The task now is to transform $g_{\mu\nu}^v$ to retarded coordinates. As explained in section \ref{ssec:matching}, there can in general not be a unique way to match the advanced and retarded supertranslation fields. However, a natural choice in a static metric is to require that the spherical metrics match: $g_{AB}^v(f^-) = g_{AB}^u(f^+)$. Therefore, we use the diffeomorphism $\mathcal{D}_m$ defined by
\begin{equation}
	v = u + 2 \limitint_{r_0}^r \frac{1}{1-\frac{2GM}{r}}  \ldiff{r'} - \frac{D^2 f^-}{1-\frac{2GM}{r}} - 2 f^- \,.
	\label{matchingDiffeomorphism}
\end{equation}
Then it turns out that
\begin{equation}
	\mathcal{D}_m\left(g_{\mu\nu}^v(f^-)\right) = g_{\mu\nu}^{u\,,0} - \mathcal{L}_{\xi_u(f^-)} g_{\mu\nu}^{u\,,0} = g_{\mu\nu}^u(-f^-)  \,.
\end{equation}
Thus, we identify
\begin{equation}
	f^+=-f^- \,.
	\label{matchingPointwise}
\end{equation}
Up to a sign, the supertranslation field in advanced coordinates matches the retarded one angle-wise. With this choice, not only the spherical metrics match, but also the $g_{00}$-components, \ie the Newtonian potentials.

\section{Explicit Solution for Goldstone Supertranslation of a Planet}
\label{app:planetSolution}

\subsubsection*{Step 1: Absorption}
The Goldstone supertranslation consists of two steps: First, an initially spherically symmetric planet absorbs as wave. As is well-known (see \eg (9.3) in \cite{hobson}), the metric of a static spherically symmetric spacetime can be cast in the general form
\begin{align}
\mathrm{d} s^2 = - A(r)\diff t^2 + B(r) \diff r^2 +r^2 \diff \Omega^2 \,, \label{generalMetric}
\end{align} 
where all physical information is contained in the $tt$- and $rr$-components. Since we want to describe a planet, there should neither be a surface of infinite redshift, \ie $A(r) > 0 \ \forall \ r$, nor an event horizon, \ie $B(r) < \infty \ \forall \ r$. Furthermore, asymptotic flatness implies that $A(r) \stackrel{r\rightarrow \infty}{\longrightarrow} 1$ and $B(r) \stackrel{r\rightarrow \infty}{\longrightarrow} 1$ sufficiently fast. Using the transformation
 \begin{align}
v = t + \limitint_{r_0}^r \diff r' \sqrt{\frac{B}{A}} \,, \label{advancedTime}
\end{align}
we obtain the metric $g_{\mu\nu}^v$ in advanced BMS-gauge:
\begin{align}
\diff s^2 = - A \diff v^2 + 2 \sqrt{AB} \diff v \diff r + r^2 \diff \Omega^2 \,, \label{planetAdvanced}
\end{align}
which is suited to describe incoming radiation. Note that this metric describes the whole spacetime and not only its asymptotic region, \ie $r\rightarrow \infty$.

We will restrict ourselves to infinitesimal supertranslations. In advanced time, these are generated by
\begin{subequations} \label{vectorFieldV}
	\begin{align}
	\xi_v^v =& f^- \,,  \\ 
	\xi_v^r =& -\frac{1}{2} r D_B \xi_v^B \,, \\
	\xi_v^A =& f^{-,\,A} \limitint_r^\infty \diff r' (\sqrt{AB} r'^{-2}) \,,
	\end{align}
\end{subequations}
where an arbitrary function $f^-$ on the sphere determines the change of the supertranslation field.  We denote it by $f^-$ instead of $T^-$ in this appendix to avoid confusion with the energy-momentum-tensor of the wave. The minus-superscript indicates that we deal with a supertranslation in advanced coordinates. Our goal is to realize the infinitesimal diffeomorphism defined by \eqref{vectorFieldV} in a physical process, \ie outside the planet, we want to have the stationary metric $g^v_{\mu\nu}$ before some time $v_0$ and after some point of time $v_1$, we want to end up in the stationary metric $g^v_{\mu\nu} + \mathcal{L}_{\xi_v(f^-)} g_{\mu\nu}^v$. For $v_0<v<v_1$, physical radiation interpolates between the two metrics. Inside the planet, the wave should be absorbed so that the transformation fades out and the metric around the origin remains unchanged. Adding as final ingredient a change of the Bondi mass $\mu$, which is necessary to ensure the positive energy condition, we obtain
\begin{align}
\delta g^v_{\mu\nu} = \tau_{v_0,\,v_1}(v) s^-(r) \left(\mathcal{L}_{\xi_v(f^-)} g^v_{\mu\nu} + \frac{2\mu G}{r} \delta_\mu^0 \delta_\nu^0 \right) \,, \label{dynamicalTransformationV}
\end{align}
where $0\leq \tau_{v_0,\,v_1}(v) \leq 1$ parameterizes the interpolation, \ie $\tau_{v_0,\,v_1}(v<v_0)=0$ and $\tau_{v_0,\,v_1}(v>v_1)=1$. The function $s^-(r)$ describes the absorption of the wave by the planet. It has the property that it is monotonically increasing with $s^-(0)=0$ and $s^-(\infty)=1$, where $s^-(0)=0$ ensures that the wave is fully absorbed before the origin and no black hole forms. Moreover, $s^{-\,'}(r)\neq 0$ is only permissible whenever the local energy density of the planet is non-zero. The magnitude of $s^{-\,'}(r)$ determines how much absorption happens at $r$. It is crucial to note that the transformation $s^-(r) \mathcal{L}_{\xi_v(f^-)} g_{\mu\nu} $ is a diffeomorphism only where $s^-(r)$ is constant. Thus, the transformation \eqref{dynamicalTransformationV} acts as a diffeomorphism only outside the planet, but not inside. This reflects the fact that we want to obtain a physically different planet. A transformation which acts as a diffeomorphism everywhere could not achieve this.

Since we work with infinitesimal supertranslations, it is important that we stay within the regime of validity of this first-order approximation, \ie that terms linear in $f^-$ dominate. As it will turn out in the calculation, this is the case if $\max_{(\theta, \varphi)} |f^-| \ll v_1-v_0$. This means that the time-shift induced by the supertranslation must be much smaller than the time-scale of the process, \ie the supertranslation must be performed slowly. We will choose $v_1-v_0$ such that this is the case and so that we can neglect all higher orders in $f^-$ when we calculate the Einstein equations.

We have to show that the transformation \eqref{dynamicalTransformationV} leads to a valid solution of the Einstein equations. Thus, if we calculate the Einstein $G_{\mu\nu}$ and consequently the new energy-momentum-tensor $T_{\mu\nu}$, we have to demonstrate that this is a valid source. To this end, we have to perform two checks. First of all, it must be conserved, $T_{\mu\nu}^{\phantom{\mu\nu};\mu} = 0$. This is trivially true in our construction because of the Bianchi identity, $G_{\mu\nu}^{\phantom{\mu\nu};\mu} = 0$. Secondly, we have to show that $T_{\mu\nu}$ fulfills an appropriate energy condition. For that purpose, we first note that this perturbation only depends on the local geometry, except for $\xi^\mu_v$ and $s^-(r)$, which also depend on spacetime points at bigger radii. Thus, outside the planet, we have the same solution as in \cite{HPS2}, except for the fact the we perform our supertranslation slowly:
\begin{subequations} \label{waveSchwarzschild}
	\begin{align}
	T_{00} =& \frac{1}{4\pi r^2} \left[\mu - \frac{1}{4} D^2(D^2 + 2) f^- + \frac{3M}{2r} D^2 f^-\right] \tau'_{v_0,\,v_1}(v) \,,\\
	T_{0A} = & \frac{3M}{8\pi r^2} D_A f^- \tau'_{v_0,\,v_1}(v) \,, 
	\end{align}
\end{subequations}
where we used that there is no absorption outside the planet: $s^-(r>R)=1$.
Obviously, the energy condition is fulfilled. At this point, we remark that leaving out all subleading parts, which are proportional to $M$, would also lead to a valid wave in the metric \eqref{planetAdvanced}, i.e., $T_{\mu\nu}^{\phantom{\mu\nu};\mu}$ would also be true to all orders if one only considered the leading order of \eqref{waveSchwarzschild}. This means that we add the subleading parts to \eqref{waveSchwarzschild} not because of energy conservation but since we want to realize the transition \eqref{dynamicalTransformationV} not only to leading order in $1/r$, but to all orders.\footnote
{Of course, energy conservation relates the two subleading parts of $T_{00}$ and $T_{0A}$. When we choose one, it determines the other.}

Fortunately, we do not either have to worry about the energy condition inside the planet. For a small enough perturbation, this is true since the energy condition inside a planet is not only marginally fulfilled. This means that $s^{-'}(r)$ can be non-zero inside the planet: This corresponds to absorption of the wave by the planet.

Lastly, we have to show that the wave is still a valid solution after it has been partly absorbed. For the purpose of illustration, we model the planet as a sequence of massive shell with vacuum in between, $T_{\mu\nu}=0$. In that case, the only non-trivial question is whether \eqref{dynamicalTransformationV} fulfills the energy condition after it passed some or all of the shells. Therefore, we calculate the energy-momentum-tensor in this region. By Birkhoff's theorem, the local geometry corresponds to a Schwarzschild solution with diminished mass $\tilde{M}$ (where $\tilde{M}$ can be zero). It only depends on the matter which it has passed via $\xi_v^\mu$. We parameterize the difference of $\xi_v^\mu$ and the vector field one would get in a pure Schwarzschild geometry of mass $\tilde{M}$ by 
\begin{align}
\sigma := \limitint_{R_{\text{min}}}^{R_{\text{max}}} \diff r' ((\sqrt{AB}-1) r'^{-2})\,,
\end{align}
where  we have no matter for $r>R_{\text{max}}$ and between $r$ and $R_{\text{min}}$.\footnote
{We use that $AB=1$ in a Schwarzschild geometry of arbitrary mass.}
Explicitly, this means that we can write
\begin{align*}
\xi_v^A =& f^{-,\,A}\left(\frac{1}{r} + \sigma\right) \,,
\end{align*}
where it is important that $\sigma$ does not depend on $r$ in our region of interest. Of course, $\sigma=0$ corresponds to the case when there is no matter outside.

With the help of Mathematica \cite{mathematica}, we compute:
\begin{subequations} \label{wavePlanetV}
	\begin{align}
	T_{00} =& \frac{1}{4\pi r^2} \left[\tilde{\mu} - \left(1+\sigma r\right)\left(\frac{1}{4} D^2(D^2 + 2) \tilde{f}^- - \frac{3\tilde{M}}{2r} D^2 \tilde{f}^-\right)\right] \tau'_{v_0,\,v_1}(v) \,,\\
	T_{0A} = & \left[\frac{3\tilde{M}}{8\pi r^2} D_A \tilde{f}^- +\frac{\sigma}{16 \pi} D_A (D^2 + 2)\tilde{f}^-\right] \tau'_{v_0,\,v_1}(v) \,,\\
	T_{AB} =& -\frac{\sigma r}{8 \pi}\left[\left(2D_A D_B - \gamma_{AB}D^2 \right)\tilde{f}^- \right]\tau'_{v_0,\,v_1}(v) \,,
	\end{align}
\end{subequations}
where $\tilde{f}^-=s^-(r)f^-$ is the supertranslation which is attenuated because of absorption in the outer shells. It is crucial to note that $s^{-\,'}(r)=0$ in this calculation since we are not inside one of the shells of the planet and likewise $\tilde{\mu} =s^-(r)\mu$.
As we can estimate $\sigma$ very crudely as $\sigma < 1/R$, we see that for sufficiently large $\mu$, the energy condition is fulfilled. With a more accurate estimate, we expect that the freedom of choosing $\mu$ is not restricted when the wave passes a massive shell. In summary, we have shown that the metric \eqref{dynamicalTransformationV}, which describes the dynamical transition from a spherically symmetric planet to a counterpart with nontrivial angular distribution of mass, is a valid solution.

\subsubsection*{Step 2: Emission}
The second step is to describe the emission of the wave by the planet. Thus, our initial metric is the one after absorption, as determined by equation \eqref{dynamicalTransformationV}:
\begin{equation}
\delta g^v_{\mu\nu} = s^-(r) \left(\mathcal{L}_{\xi_v(f^-)} g^v_{\mu\nu} + \frac{2\mu G}{r} \delta_\mu^0 \delta_\nu^0 \right) \,.
\label{advancedMetricStatic}
\end{equation}
As we want to consider emission, our first step is to transform it to retarded coordinates. Intuitively, it is clear that it should be possible to describe a slightly asymmetrical planet also in retarded coordinates. While it is generically hard to write down the corresponding diffeomorphism which connects the two metrics, we can use that the metric of a planet does not differ from Schwarzschild in the exterior region. Therefore, we can use the diffeomorphism \eqref{matchingDiffeomorphism} to obtain
\begin{equation}
g_{\mu\nu}^u = 	g_{\mu\nu}^{u,\,0} + s^-(r)\left(\mathcal{L}_{\xi_u(-f^-)} g_{\mu\nu}^{u,\,0} + \theta(R-r)\text{dev} \right) \,,
\label{retardedMetricTranslated}
\end{equation}
where $g_{\mu\nu}^{u,\,0}$ is the metric of the initial, spherically symmetric planet in retarded coordinates. This means that we apply a supertranslation in retarded coordinates which is defined by the function $f^-$ used to defined the advanced supertranslation. The function $\text{dev}$ accounts for the fact that we do not know the continuation of the matching diffeomorphism \eqref{matchingDiffeomorphism} to the interior of the planet. Therefore, $g_{\mu\nu}^u$ might deviate slightly from BMS-gauge but only in the interior. We expect, however, that the matching diffeomorphism can be continued such that $\text{dev}=0$. Finally, we want to point out that $g_{\mu\nu}^u(r=0) = g_{\mu\nu}^{u,\,0}(r=0)$ since $s^-(r=0)=0$, \ie the wave does not reach the center and the mass distribution of the planet is still spherically symmetric around $r=0$.

The case of the planet provides us with another justification why the matching \eqref{matchingPointwise} is natural. With this identification, both the metric \eqref{advancedMetricStatic} in advanced coordinates and the metric \eqref{retardedMetricTranslated} in retarded coordinates cover the whole manifold. Extrapolating the results of \cite{compereMatching, compereFinite}, where finite supertranslations of Schwarzschild and Minkowski are discussed, we expect that for any other matching, \ie for any other value of the supertranslation field, this is no longer the case. If this is true, the requirement that the BMS-coordinate system covers the whole manifold singles out a unique value of the advanced supertranslation field as well as a unique value of the retarded supertranslation field, and therefore a coordinate matching.

Next, we want to describe how the metric \eqref{retardedMetricTranslated} emits a wave. This wave should realize a supertranslation described by $f^+$, which is generically different from $f^-$:
\begin{align}
\delta g^u_{\mu\nu} = \tau_{u_0,\,u_1}(u)s^+(r) \left(\mathcal{L}_{\xi_u(f^+)} g^{u,\,0}_{\mu\nu} - \frac{2\mu G}{r} \delta_\mu^0 \delta_\nu^0 \right) \,, \label{dynamicalTransformation2}
\end{align}
where we used that $\mathcal{L}_{\xi_u(f^+)} g^{u}_{\mu\nu} = \mathcal{L}_{\xi_u(f^+)} g^{u,\,0}_{\mu\nu}$ to first order in $f^+$ and $f^-$. Thus, working only to first order simplifies our calculations significantly since we can simply use the calculations for the absorption. The wave \eqref{wavePlanetV} becomes:
\begin{subequations} \label{wavePlanetU}
	\begin{align}
	T_{00} =& \frac{1}{4\pi r^2} \left[\tilde{\mu} - \left(1+\sigma r\right)\left(\frac{1}{4} D^2(D^2 + 2) \tilde{f}^+ - \frac{3\tilde{M}}{2r} D^2 \tilde{f}^+\right)\right] \tau'_{u_0,\,u_1}(u) \,,\\
	T_{0A} = & -\left[\frac{3\tilde{M}}{8\pi r^2} D_A \tilde{f}^+ +\frac{\sigma}{16 \pi} D_A (D^2 + 2)\tilde{f}^+\right] \tau'_{u_0,\,u_1}(u) \,,\\
	T_{AB} =& -\frac{\sigma r}{8 \pi}\left[\left(2D_A D_B - \gamma_{AB}D^2 \right)\tilde{f}^+ \right]\tau'_{u_0,\,u_1}(u) \,.
	\end{align}
\end{subequations}
As for the absorption, we have shown that we can realize the transformation \eqref{dynamicalTransformation2} with a physical wave.

Finally, we analyze the joint effect of absorption and emission. Combining the transformations \eqref{dynamicalTransformationV} and \eqref{dynamicalTransformation2}, we get total total change of the metric:
\begin{align}
	\delta g_{\mu\nu}^{\text{tot}}& = \theta(r-R) \mathcal{L}_{\xi_u(f^+-f^-)} g_{\mu\nu}^{u,\,0} \nonumber\\
	& + \theta(R-r)\left(s^+(r)\mathcal{L}_{\xi_u(f^+)} g_{\mu\nu}^{u,\,0} - s^-(r) \mathcal{L}_{\xi_u(f^-)} g_{\mu\nu}^{u,\,0} + \text{dev}  \right) \,,
	\label{totalTransformation}
	\end{align}
	where we used retarded coordinates. As desired, the mass of the planet stays invariant. Moreover, $\delta g_{\mu\nu}^{\text{tot}}$ acts as a diffeomorphism outside the planet, namely it is the difference of the advanced supertranslation, described by $f^-$, and the retarded supertranslation, described by $f^+$. If we furthermore assume that the term $\text{dev}$, which reflects our incomplete knowledge of the matching between advanced and retarded coordinates in the planet, is zero, we see that the metric does not change for $f^-=f^+$. We obtain a trivial transformation if the angular energy distribution of ingoing and outgoing radiation is angle-wise the same.

 \section*{Acknowledgements}

We are happy to thank Gia Dvali for many stimulating discussions and comments. We also thank Artem Averin, Henk Bart and Raoul Letschka for discussions. The work of C.G. was supported in part by Humboldt Foundation and by Grants: FPA 2009-07908 and ERC Advanced Grant 339169 "Selfcompletion".

\addcontentsline{toc}{section}{References}

\providecommand{\href}[2]{#2}\begingroup\raggedright\endgroup

\end{document}